%% file: ms.tex
\documentclass[aps,pra,reprint,groupedaddress,showemail,showpacs,nofootinbib]{revtex4-1}

\usepackage{amsthm}
\usepackage{amsmath}
\usepackage{latexsym}
\usepackage{amsfonts}
\usepackage{amssymb}
\usepackage{color}
\usepackage{bbm,dsfont}
\usepackage{graphicx}
\usepackage{enumerate}
\usepackage{hyperref}
\usepackage{subfigure} 
\usepackage{tikz}
\usepackage{placeins}
\usepackage{import}
\usepackage{enumitem}
\usepackage[utf8]{inputenc}
\newcommand\opone{\leavevmode\hbox{\small1\kern-3.8pt\normalsize1}}

\DeclareUnicodeCharacter{00A0}{ }

\usepackage{bbold}

\def\W{\operatorname{W}}
\def\U{\operatorname{U}}

\def\Tr{\operatorname{Tr}}
\def\T{\operatorname{T}}
\def\Q{\operatorname{Q}}

\def\sigmap{\operatorname{\sigma_+}}
\def\sigmam{\operatorname{\sigma_-}}
\def\sigmax{\operatorname{\sigma_x}}
\def\sigmay{\operatorname{\sigma_y}}
\def\sigmaz{\operatorname{\sigma_z}}

\newtheorem{proposition?}{Proposition?}

\theoremstyle{definition}

\newcommand{\real}{\mathbb R} 

\newcommand{\hi}{\mathcal{H}} 
\newcommand{\ket}[1]{|#1\rangle} 
\newcommand{\bra}[1]{\langle#1|} 
\newcommand{\kb}[2]{|#1\rangle\langle#2|} 

\newcommand{\id}{\mathbbm{1}} 

\newcommand{\vc}{\mathbf{c}} 
\newcommand{\vn}{\mathbf{n}} 
\renewcommand{\vr}{\mathbf{r}} 
\newcommand{\vm}{\mathbf{m}} 
\newcommand{\vk}{\mathbf{k}} 
\newcommand{\vx}{\mathbf{x}} 
\newcommand{\vy}{\mathbf{y}} 
\newcommand{\vz}{\mathbf{z}} 
\newcommand{\vsigma}{\boldsymbol{\sigma}} 

\newcommand{\matt}[1]{\left( \begin{array}{cc} #1 \end{array} \right)} 

\newcommand{\E}{\mathsf{E}}

\newcommand{\So}{\mathsf{S}}

\newcommand{\Ech}{\mathcal{E}} 

\newcommand{\Sys}{\mathcal{S}} 
\newcommand{\Anc}{\mathcal{A}} 
\newcommand{\Cont}{\mathcal{C}} 

\begin{document}


\title{Collision model approach to steering of an open driven qubit}

\author{Konstantin Beyer} \email{konstantin.beyer@tu-dresden.de}
\affiliation{Institut f{\"u}r Theoretische Physik, Technische
  Universit{\"a}t Dresden, D-01062, Dresden, Germany}

\author{Kimmo Luoma} \email{kimmo.luoma@tu-dresden.de}
\affiliation{Institut f{\"u}r Theoretische Physik, Technische
  Universit{\"a}t Dresden, D-01062, Dresden, Germany} \author{Walter
  T. Strunz} \email{walter.strunz@tu-dresden.de} \affiliation{Institut
  f{\"u}r Theoretische Physik, Technische Universit{\"a}t Dresden,
  D-01062, Dresden, Germany}

\date{\today}

\begin{abstract}
  We investigate quantum steering of an open quantum system by
  measurements on its environment in the framework of collision
  models.  As an example we consider a coherently driven qubit
  dissipatively coupled to a bath. We construct local non-adaptive and
  adaptive as well as nonlocal measurement scenarios specifying
  explicitly the measured observable on the environment.  Our approach
  shows transparently how the conditional evolution of the open system
  depends on the type of the measurement scenario and the measured
  observables. These can then be optimized for steering.  The nonlocal
  measurement scenario leads to maximal violation of the used steering
  inequality at zero temperature.  Further, we investigate the
  robustness of the constructed scenarios against thermal noise.  We
  find generally that steering becomes harder at higher temperatures.
  Surprisingly, the system can be steered even when bipartite
  entanglement between the system and individual subenvironments
  vanishes.

\end{abstract}


\maketitle

\section{Introduction}
\label{sec:introduction}
Perhaps the most fascinating feature of quantum theory is the
existence of correlations that cannot be explained by classical local
randomness. Since the dawn of the quantum age it has been known that
entanglement, a term coined by Schrödinger \cite{schroedinger}, can
give rise to paradoxes when seen in the light of a realistic theory as
pointed out by EPR~\cite{epr}. About 30 years later, a formulation
that allowed experimental testing of nonlocal phenomena of the theory
was given by Bell~\cite{Bell:1964kc}.  More recently, Wiseman {\it
  et.al.} discussed another facet of quantum correlations, namely
steerability. Indeed, they showed that nonlocality, steerability and
entanglement form a strict hierarchy~\cite{wiseman-steering}.
Steering is a task where one party is trying to remotely influence
another party's state by local measurements. Over the last decade this
phenomenon has been under intensive theoretical
\cite{jones,steering-criteria,
  costa-steering,costa-channels,nguyen-geometry,PhysRevLett.115.230402,
  PhysRevLett.113.160403,PhysRevLett.112.180404,PhysRevLett.113.160402}
and experimental
\cite{saunders2010,wittman2012,PhysRevLett.116.160403,
  PhysRevLett.118.140404} investigations.

Mostly, steering has been analyzed in situations where quantum
dynamics plays no role. In general, quantum systems are coupled to
their environment and thus they are open.  Let us assume that the open
system and the environment both are initially in a pure
state. Accordingly, the joint state then evolves as
\begin{equation}
  \label{eq:unitary}
  \ket{\Psi_t} = \U_t (\ket{\psi_\Sys} \otimes \ket{\psi_\Anc}),
\end{equation}
where $\U_t$ is a unitary transformation depending on time $t$ and
$\ket{\psi_\Sys}$ and $\ket{\psi_\Anc}$ are the initial states of the
system and the environment, respectively. {The reduced system state is
  then given by $\rho_\Sys=\Tr_\Anc[\ket{\Psi_t}\bra{\Psi_t}]$, where
  $\Tr_\Anc$ denotes the partial trace over the environment.  While
  the joint state remains pure for all times, the reduced state
  becomes mixed due to interactions giving rise to entanglement
  creation
  \begin{equation}
    \Tr[\rho_\Sys^2] < 1 \iff \ket{\Psi_t} \text{ is entangled.}
  \end{equation}}

Pure entangled states can always be used for quantum steering tasks
\cite{gisin,wiseman-steering}. Therefore, one may ask whether and how
the system can be steered by measurements on its environment.  Based
on this idea, an experiment for showing that quantum jumps in a
coherently driven two-level atom are detector dependent was proposed
{and investigated}
in~\cite{wiseman-gambetta,daryanoosh_quantum_2014,daryanoosh_detector_2015}.
The system under consideration there can be described by the
Gorini-Kossakowski-Sudarshan-Lindblad (GKSL) master equation for
resonance fluorescence
\cite{GoriniKossakowskiSudarshan,lindblad,carmichael}
\begin{equation}
  \label{eq:gksl}
  \dot{\rho}_\Sys = \mathcal{L}\rho_\Sys = 
  -i\omega[\sigmax,\rho_\Sys] 
  + \gamma \sigmam \rho_\Sys \sigmap 
  - \frac{\gamma}{2}\{\rho_\Sys,\sigmap \sigmam\},
\end{equation}
where $\omega$ is the driving strength and $\gamma$ is the damping
rate. Master equations, such as Eq.~(\ref{eq:gksl}), can be unraveled
in many different ways producing ensembles that all represent the same
reduced state $\rho_\Sys$~\cite{carmichael,wiseman-milburn,robust}.
{In~\cite{wiseman-gambetta,daryanoosh_quantum_2014,daryanoosh_detector_2015}}
different adaptive and non-adaptive photo detection schemes are
implemented, corresponding to different unravelings, and it was shown
that the produced ensembles are able to violate a steering inequality.

We approach the task of steering an open quantum system using
collision models~\cite{ziman-collision,ziman-decoherences,
  ziman-diluting,scarani-thermalizing,
  giovannetti-collision,giovannetti-correlated,filippov2017,composite-collision}.
In unravelings, the continuous monitoring of the environment results
in a stochastic evolution equation describing the open system
dynamics.  In the framework of collision models, {the bath consists of
  many subenvironments interacting discretely and individually with
  the open quantum system. The measurements on the environment can
  thus be implemented explicitly as a sequence of measurements on the
  subenvironments.} In this Article we construct suitable local and
nonlocal measurement scenarios on the environment, specifying the
measured observables, to achieve steering of an open system described
by Eq.~(\ref{eq:gksl}) in the time-continuous limit.  Further, we
extend the discussion to the case where the open system is coupled to
a thermal bath.

The Article is organized as follows. In Sec.~\ref{sec:concepts} we
briefly review the concepts of collision models, discrete quantum
trajectories, measurement scenarios and quantum steering.  In
Sec.~\ref{sec:example} we apply these concepts to the model system. We
show that a {monitored} collision model is suitable for the
description of steering in such a system. Specifically, in
Sec.~\ref{sec:nonlocal} we provide a concrete realization of a
nonlocal measurement scenario, which in the time-continuous limit
leads to a Markovian two-qubit GKSL equation, embedding
Eq.~(\ref{eq:gksl}).  In Sec.~\ref{sec:thermal} we discuss the thermal
case and show that steering is possible even when bipartite
entanglement between the system an individual subenvironments vanishes. Lastly, in Sec.~\ref{sec:conclusion} we conclude.

\section{Concepts}
\label{sec:concepts}
\subsection{Collision model}
\label{sec:cm}
\begin{figure}[ht]
  \def\svgwidth{\columnwidth} 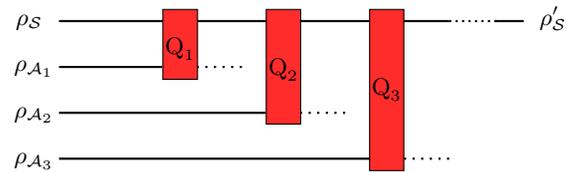
  \caption{{Scheme of a collision model with the first three
      collisions between $\Sys$ and the subenvironments $\Anc_i$. The
      subenvironments are traced out after their
      collisions.}\label{fig:cm}}
\end{figure}

Let $\rho_\Sys$ be the density operator describing the open quantum
system $\Sys$ with Hilbert space $\hi_\Sys$. {The environment $\Anc$
  of the open system is composed of subenvironments $\Anc_i$ with
  Hilbert spaces $\hi_{\Anc_i}$. In the class of collision models we
  consider, the open system interacts sequentially and only once with
  each subenvironment.}  {The coupling between $\Sys$ and $\Anc_i$ is
  given by a unitary {operator} $\Q_i$. The reduced system state after
  $n$ collisions then reads \cite{ziman-collision}}
\begin{equation}
  \rho_\Sys(n) = \Tr_{\Anc}[\Q_n \dots \Q_1(\rho_S(0) \bigotimes_{i=1}^n \rho_{\Anc_i})\Q_1^{\dagger} \dots 
  \Q_n^{\dagger}],
  \label{eq:cm}
\end{equation} 
{where $\rho_\Sys(0)$  and
  $\bigotimes_{i=1}^n \rho_{\Anc_i}$ are the initial states of open system
  and environment, respectively.} We
restrict ourselves here to the case where the subenvironments are
initially in a product state. Furthermore, we assume that each unitary
$\Q_i$ acts {nontrivially} only on the system $\Sys$ and 
the $i$th subenvironment
$\Anc_i$. {Therefore, it is possible to trace out the $i$th 
  subenvironment $\Anc_i$ 
  right after the $i$th collision,  without changing the future evolution of
  the reduced state and  we can write recursively} 
\footnote{Please note that these restrictions are in general not
  necessary for collision models. Initial entanglement in the environment or
  couplings between the subenvironments can lead to memory effects and
  are useful in order to model, for example, non-Markovian
  dynamics. We refer to~\cite{memory-kernel,kretschmer,
    luoma-nm,rybar,ciccarello-nm,cakmak-collision}
  for further details.} 
\begin{equation}
  \rho_\Sys(i) = \Tr_{\Anc_i}[\Q_i(\rho_\Sys(i-1) \otimes \rho_{\Anc_i})\Q_i^{\dagger}].
  \label{eq:cm-single-step}
\end{equation}
{We clearly see that the $i$th state $\rho_\Sys(i)$ of the 
  open system depends only on $\Q_i,\, \rho_\Sys(i-1)$ and 
  $\rho_{\Anc_i}$ and not on earlier evolution of the open system.
  Hence, the evolution of the open 
  system is memoryless.}
{The collision model discussed in our work is shown schematically in Fig.~\ref{fig:cm}.}

\subsection{Discrete quantum trajectories}

Measuring selectively the subenvironments after their interaction with
the system --- instead of tracing them out --- leads to a discrete
conditional evolution for $\Sys$~\cite{brun}. {While the evolution of
  the reduced state $\rho_\Sys$ is deterministic, the evolution of
  conditional states is stochastic, due to the probabilistic nature of
  quantum measurements.}

{We denote the joint state of the system and the first subenvironment
  after their collision by
  {$\rho_{\Sys\Anc_1}=\Q_1(\rho_\Sys(0)\otimes\rho_{\Anc_1})Q_1^\dagger$. }
  {The conditional state of the system $\rho_{k|E}$ after a
    measurement on the {sub}environment is~\cite{nielsen}
    \begin{equation}
      \rho_{k|E} =\frac{\Tr_{\Anc_1}\left[\rho_{\Sys\Anc_1} (\id_\Sys \otimes E_{k})\right]}
      {p(k)},
    \end{equation}
    {where $k$ labels the measurement outcomes of the positive
      operator valued measure (POVM) ${\{E_{k}\}}$ and
      $p(k)=\Tr\left[\rho_{\Sys\Anc_1} (\id_\Sys \otimes E_k)\right]$}
    {is the probability to obtain outcome $k$ when measuring
      observable $E$}.  {This process can be repeated using
      $\rho_{k|E}$ as the initial system state for the next collision
      and the subsequent measurement.}}

  {In this way a sequence of states $\{\rho_{k_i|E_i}\}_{i=1}^N$,
    called a \textit{discrete quantum trajectory}, is obtained.  Each
    $\rho_{k_i|E_i}$ is conditioned on all measurements and their
    respective outcomes up to and including the $i$th step.  We call
    the last state $\rho_{k_N|E_N}$ the \textit{trajectory endpoint},
    where $N$ is the number of collisions.}

  {The reduced state after the $i$th collision $\rho_\Sys(i)$ is
    recovered by averaging the conditional states at the $i$th step over
    all possible trajectories weighted by their probability of
    occurrence.}

  \subsection{Local and nonlocal measurement
    scenarios}\label{sec:scenarios} {A set of rules, which determines
    how the subenvironments are measured, is called a
    \textit{measurement scenario}~\cite{wiseman-milburn}}.
  \paragraph*{Local scenario.}  {Let $\mathcal{X}_i$ be the set of all
    possible {POVMs} on $\hi_{\Anc_i}$.  For any
    $X_i\in\mathcal{X}_i$, let $\Omega_{X_i}$ be the set of outcomes.
    A local measurement scenario for $N$ collisions is a rule how to
    construct the {generalized observable}
    $A\in \mathcal{X}_1\times\cdots\times\mathcal{X}_N$. Further, the
    choice of $A$ fixes the outcome space
    $\Omega_A=\Omega_{X_1}\times\cdots\times\Omega_{X_N}$.}
  We call a local measurement scenario \textit{non-adaptive} if all
  {subenvironments are measured locally and independently of each
    other}. Local scenarios appear as the natural choice for the
  step-by-step structure of a collision model. Each {subenvironment}
  may be measured right after its interaction with the system and can
  be discarded afterwards (see Fig.~\ref{fig:cm-measurement}). In a
  local non-adaptive measurement scenario, the measured observable $A$
  can be predefined before the start of the collision dynamics.
  \begin{figure}[ht]
    \def\svgwidth{\columnwidth}
    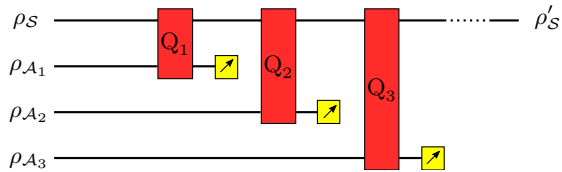
    \caption{Scheme of a collision model with local measurements.  The
      open system starts in the state $\rho_\Sys$ and interacts
      sequentially with the {subenvironments} which are initially in
      the states $\rho_{\Anc_i}$. The interaction is described by
      unitary transformations $\Q_i$ and the {subenvironments} are
      measured after the interaction.}\label{fig:cm-measurement}
  \end{figure}

  We can think of a local scenario where the choice of the measurement
  on the $i$th subenvironment $\Anc_i$ may depend on previously
  measured observables {$\{X_j\}_{j=1}^{i-1}$} and on their outcomes
  $\{k_j\}_{j=1}^{i-1}$.  A scenario which makes use of this classical
  information in order to choose the measurement for the next step is
  called an \textit{adaptive} measurement
  scenario~\cite{wiseman_quantum_1999,wiseman-milburn,karasik_how_2011}}.
In a local adaptive measurement scenario, the measured observable $A$
is determined during runtime.
\paragraph*{Nonlocal scenario.}
\label{sec:nonlocal-scenarios}
Even though local measurements are directly motivated by the structure
of collision models, it is worth {considering} more general
measurements. After $N$ collisions without measurements the joint
state of the system and the environment is
\begin{equation}
  \rho_{\Sys\Anc_1\cdots\Anc_N} = \Q_N \cdots \Q_1(\rho_\Sys(0) 
  \bigotimes_{i=1}^N \rho_{\Anc_i})\Q_1^{\dagger} \cdots \Q_N^{\dagger}.
  \label{eq:joint-state}
\end{equation}
{The most general observable $A^{\textup{nonloc}}$ to be measured on the 
  environment can be any from the set of observables on 
  $\hi_{\Anc_1}\otimes\cdots\otimes\hi_{\Anc_N}$ and is generally 
  nonlocal.
  In order to measure such a general observable, all of 
  the subenvironments  have to be stored after
  their interaction with the system. Accordingly, we lose the 
  advantages of handling only a few subenvironments instead of
  the whole environment.}
{For this reason we now introduce a class of  nonlocal
  measurements which is motivated by the structure of collision models
  and circumvents the storage of many subenvironments.}

\begin{figure}[ht]
  \def\svgwidth{\columnwidth} 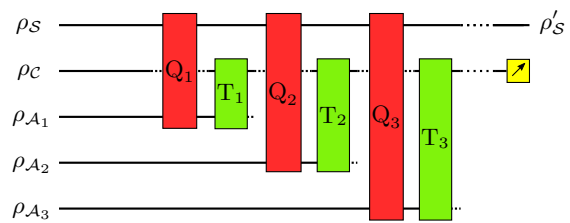
  \caption{{A collision model {which is equivalent to} a nonlocal
      measurement on all involved subenvironments.  After its coupling
      with the system, each subenvironment interacts with the control
      system by a unitary transformation $\T_i$. There is only one
      measurement in the end performed on the control system
      $\Cont$.}}\label{fig:cm-control}
\end{figure}

The measurement scenario is schematically shown in
Fig.~\ref{fig:cm-control}. Instead of being measured after their
interaction with the system, the subenvironments are coupled to an
ancilla system $\Cont$, called the \textit{control system}, which does
not belong to the environment but is a part of the measurement
apparatus. The interaction is mediated { by unitary} transformations
$\T_i$ acting {nontrivially} on the control system and the respective
subenvironment $\mathcal{A}_i$. After the $\T$-gate the
{subenvironments} may be traced out. {{The actual measurement is
    performed on {$\Cont$} after a sequence of $N$ collisions and the
    {POVM} $C=\{C_k\}$ can be any from the set of {POVMs} on
    $\hi_\Cont$.} If the {operators $\T_i$ are nonlocal unitaries},
  then the measurement on $\Cont$ is in general equivalent to some
  nonlocal measurement on all subenvironments. For a certain {POVM
    element} $C_k$ we have
  \begin{equation}
    p_k = \Tr[\mathcal{T_{\Anc\Cont}}(\rho_{\Sys\Anc_1\cdots\Anc_N}\otimes\rho_\Cont)\mathcal{T_{\Anc\Cont}}^\dagger(\id_{\Sys\Anc_{1}\ldots \Anc_{N}}\otimes C_k)],
  \end{equation}
  where $\mathcal{T}= \T_N \circ \ldots \circ \T_1$. Therefore, due to
  the cyclic property of the trace, the nonlocal observable measured
  by the given measurement scenario is
  \begin{equation}
    A_k=\Tr_\Cont[\mathcal{T}_{\Anc\Cont}^\dagger(\id_{\Anc_{1}\ldots \Anc_{N}}\otimes C_k)\mathcal{T}_{\Anc\Cont}].
  \end{equation}}

{It has to be emphasized that in all of these collision models
  measurements on the environment --- no matter to which scenario they
  may belong --- do not change the reduced dynamics of the system
  described by Eq.~(\ref{eq:cm-single-step}). This can be most easily
  understood from
  Figs.~\ref{fig:cm-measurement}~and~\ref{fig:cm-control}, where we
  see that the measurements as well as the couplings $\T_i$ to the
  control system take place \textit{after} the interactions $\Q_i$
  between the subenvironments and the open system. Since $\Sys$,
  $\Anc_i$'s and $\Cont$ are all initially uncorrelated, the reduced
  dynamics of $\Sys$ remains unaffected.}

{An experimenter who has only access to the system but not to the
  environment will never be able to determine how the environment is
  measured.}

\subsection{Quantum steering}
\label{sec:steering} {Conditional dynamics of a collision model can be
  seen as a multipartite steering task.} The open system is steered by
the measurements on the {subenvironments}.
\cite{wiseman-steering,wiseman-gambetta}. For local scenarios the
measurements on the first $N$ subenvironments steer the system to the
trajectory endpoint. In the nonlocal case the steering is performed by
measuring the control system after $N$ {collisions}.

{To formulate the steering scenario, we introduce Alice and Bob.} The
latter has access to the open system $\mathcal{S}$. He can perform
arbitrary measurements on his part. Alice cannot directly act on Bob's
system but she can perform measurements on the environment, that is on
the $\mathcal{A}_i$'s. Note that in the case of a non-adaptive
scenario, Alice can measure all subenvironments at once {(possibly
  nonlocally)} and {conditional dynamics} can also be seen as a
bipartite steering task. If the measurements on the subenvironments
are adaptive, then the scenario corresponds to a multipartite steering
task with classical communication {between the parties trying to
  steer}.

{{An important} characteristic of any measurement scenario is the set
  of possible endpoints which the scenario is able to produce. We call
  this set of states the \textit{endpoint ensemble}.} A single run of
the experiment of course only produces a single trajectory with a
single (random) endpoint. The endpoint ensemble consists of all
possible trajectory endpoints which can be reached using a given
measurement scenario. If Alice finds two scenarios which lead to
different endpoint ensembles, we have steering in the sense {pointed
  out by Schr\"odinger already} in 1935~\cite{schroedinger}.

{A scenario is said to be \textit{steerable}} if the joint probability
distribution for Alice's and Bob's measurement outcomes
\textit{cannot} be written as~\cite{wiseman-steering,jones}:
\begin{equation}
  P(a,b|A,B) = \sum_{\xi} p(\xi) P_{\Anc}(a|A;\xi)P_{\Sys}(b|B;\rho_\xi),
  \label{eq:probabilities-steering}
\end{equation}  
where $A$ and $B$ are the measurements which Alice and Bob perform on
their respective parts, $a$ and $b$ are the outcomes, $p$ is a
probability distribution and $\xi$ is a local hidden variable.
$P_{\Anc}$ is a local probability distribution which can be arbitrary
because Bob does not trust Alice, so she could announce outcomes
according to an arbitrary distribution. By contrast, the distribution
at Bob's side $P_{\Sys}$ is a quantum probability distribution because
Bob trusts himself that he indeed performs quantum measurements
\cite{jones}.

{Next we discuss how steering is defined in collision models, thus
  specifying what the observables $A$, $B$ and their outcomes $a$, $b$
  are.  Alice wants to steer the state of the open system after $N$
  collisions. Therefore we are interested {in} the steerability
  properties of the joint state $\rho_{S\Anc_1\cdots\Anc_N}$ produced
  by the collision dynamics.}

In the local adaptive and non-adaptive measurement scenarios the
outcomes $a$ of Alice's observable $A$ are random vectors of length
$N$ on the outcome space $\Omega_A$.

The restricted class of nonlocal measurement scenarios is equivalent
to the usual bipartite steering scenarios, since Alice measures a
single observable $A$ on the control system.

It is in general very difficult to proof that a measured probability
distribution cannot be decomposed as in
Eq.~(\ref{eq:probabilities-steering}).  In practice, steering is
witnessed {by using} inequalities which are violated {only if} Alice
is able to steer Bob's system. There are several different forms of
such inequalities (see
e.g.~\cite{wiseman-steering,jones,steering-criteria}), depending on
the concrete systems involved.

\section{Steering of a Driven Qubit}
\subsection{The model}
\label{sec:example} {In this section we will apply the different types
  of measurement scenarios to a resonantly driven two-level system
  coupled to a vacuum bath.} We will see that the different
measurement scenarios lead to different endpoint ensembles and,
furthermore, that these ensembles are indeed able to violate a
steering inequality.

The system of interest is a {two-level system}.  {The environment
  consists of qubit subenvironments, initially in a product state
  $\rho_{\Anc}=\bigotimes_i\kb{0}{0}$, where $\ket{0}$ is the ground
  state.} {The interaction between the system and a single
  subenvironment is given by the unitary transformation}
\begin{equation}
  \W_i = e^{-i g (\sigmap \otimes \sigmam^i + \sigmam \otimes \sigmap^i)},
\end{equation}
where $g$ is a real coupling constant~\cite{brun}. The driving is
modeled by {a} local unitary transformation only acting on $\rho_\Sys$
\begin{equation}
  \U = e^{-i f \sigmax},
\end{equation}
where $f$ is a real constant. { We set $\Q_i =\W_i\U$, and define a
  single step map $\Ech_i:\rho_\Sys\mapsto \Ech_i(\rho_\Sys)$ for the
  reduced state using Eq.~(\ref{eq:cm-single-step})~\cite{kretschmer}
  \begin{equation}
    \Ech_i(\rho_\Sys) = \Tr_{\Anc_i}[\Q_i (\rho_\Sys \otimes \kb{0}{0}_i)\Q_i^\dagger].
    \label{eq:map}
  \end{equation}
  Since the collision model is homogeneous the action of the map $\Ech_i$
  is the same for any $i$ and we can set $\Ech_i = \Ech$.}

{The steady state of the map is $\Ech(\rho_{SS})=\rho_{SS}$ and its
  Bloch vector reads}
\begin{equation}
  \vr_{SS} = \begin{pmatrix}
    x_{SS}\\ 
    y_{SS} \\ 
    z_{SS}
  \end{pmatrix} = \begin{pmatrix}
    0 \\
    \frac{4 \sin ^2\left(\frac{g}{2}\right) \cos (g) \sin (2 f)}{-4 \cos (g) \cos ^2(f)+\cos (2 g)+3} \\
    \frac{(2-2 \cos (g)) (\cos (g) \cos (2 f)-1)}{-4 \cos (g) \cos ^2(f)+\cos (2 g)+3} \\
  \end{pmatrix}.
  \label{eq:steady-bloch}
\end{equation} 

It is well known that this {collision model} can reproduce the master
equation (\ref{eq:gksl}) in the time-continuous limit
\cite{brun,kretschmer}. {We define $g = \sqrt{\gamma \delta t}$ and
  $f = \omega \delta t$ and expand the map (\ref{eq:map}) to first
  order in $\delta t$. In the limit $\delta t \rightarrow 0$} we then
obtain the GKSL master equation~(\ref{eq:gksl}) {describing a damped
  driven two-level system at zero temperature~\cite{carmichael}}.

\subsection{Steering inequality} {The steering inequality which we use
  was proposed in \cite{wiseman-gambetta}. It only depends on
  quantities which can be {calculated from} the endpoint ensembles
  which Alice produces after $N$ collisions but not on the full
  sequence of outcomes.}  {The endpoint ensemble produced by a
  measurement scenario $Z$ consists of states $\rho^Z_l$, where $l$
  labels the ensemble members with
  $\sum_{l}p^Z(l)\rho_l^Z =\rho_\Sys(N)$.}  {By
  $\E_{Z}[\langle \sigma_{\vn} \rangle^2]=\sum_l p^Z(l)\langle
  \sigma_{\vn}(l)\rangle^2$, with
  $\langle \sigma_{\vn}(l)\rangle= \Tr[\rho_l^Z(\vn\cdot\vsigma)]$, we
  denote the average of the squared spin component
  $\langle \sigma_{\vn} \rangle^2$ in the direction $\vn$ over the
  ensemble produced by measurement scenario $Z$}.  {It is important to
  remember that the states of {the endpoint} ensemble depend on the
  full sequence of outcomes.}

{Quantum steering can be demonstrated if Alice is able to produce
  three endpoint ensembles $\{\rho_l^{Z_1}\}$, $\{\rho_l^{Z_2}\}$ and
  $\{\rho_l^{Z_3}\}$ which violate an inequality of the form:
  \begin{equation}
    \label{eq:inequality}
    \E_{Z_1}[\langle \sigma_{\vn} \rangle^2 ] + \E_{Z_2}[\langle \sigma_{\vm} \rangle^2 ] 
    + \E_{Z_3}[\langle \sigma_{\vk} \rangle^2 ] \leq 1,
  \end{equation}
  with $\vn\perp\vm\perp\vk$}. If Alice cannot produce three but only
two different ensembles, the inequality can still be violated by
setting $Z_2 = Z_3$.

{It is inherent in a steering task that Bob does not trust} Alice, so
we have to describe how Bob can verify that Alice indeed has produced
the ensemble she claims and how {Bob} can calculate the values needed
to check inequality (\ref{eq:inequality}).

{Alice and Bob} perform many runs of the same experiment
\cite{wiseman-gambetta}:
\begin{enumerate}
\item {Bob prepares an initial state $\rho_\Sys$ and tells Alice which
    ensemble he would like her to produce.}
\item They start the experiment and Alice {uses a measurement scenario
    that produces Bob's desired ensemble}.
\item After {$N\gg 1$} steps Bob stops the interaction with Alice's
  part and performs a measurement which he chooses randomly from a set
  of informationally complete observables~\cite{tomography,
    informationally, quorum}.
\item Alice tells Bob which trajectory endpoint she has {produced} in
  this run {by using her knowledge of the experiment and her
    measurement outcomes}.
\item Bob writes his outcome and the measurement he has performed on a
  slip of paper and throws it into a bin which is labeled by the
  endpoint which Alice has told him, and the measurement scenario she
  has used.
\item They jump back to point 1 and start the next run.
\end{enumerate}

{The number of bins Bob needs depends on how many different trajectory
  endpoints Alice announces during the runs. If the endpoint ensembles
  which Alice can produce consist of only a few points, then Bob will
  only have to label bins for these endpoints. In order to check
  whether Alice indeed produces the states she announces, Bob has to
  perform quantum tomography for each bin. Accordingly, he has to
  ensure that he has collected enough entries for each bin. The risk
  of being cheated by Alice decreases with increasing number of
  runs. When Bob has stopped the runs, he can reconstruct the endpoint
  ensembles as follows.}

\begin{enumerate}[label=\alph*)]
\item {For each bin he does quantum tomography using the information
    on the slips.}
\item {He sorts all the bins according to the measurement scenarios.}
\item {The reconstructed states belonging to the same measurement
    scenario form the endpoint ensemble of that scenario.}
\end{enumerate}

{The reconstructed ensembles should agree with the ensembles that
  Alice has reported to Bob, if she has been honest~\footnote{There
    might be a unitary rotation of all ensembles if Bob's and Alice's
    basis definitions do not agree but this is unimportant for the
    success of the steering task.}.}  {If the steering inequality is
  violated, Bob can be sure that his system has really interacted and
  has become entangled with Alice's qubits. Violation also verifies
  that Alice has not used some hidden stochastic pure state model to
  produce the trajectories.}

In general there are infinitely many possible measurement
scenarios. We will consider some examples which lead to very different
endpoint ensembles.

\subsection{Local non-adaptive scenarios}

\begin{figure}[ht]
    \def\svgwidth{.9\columnwidth}
    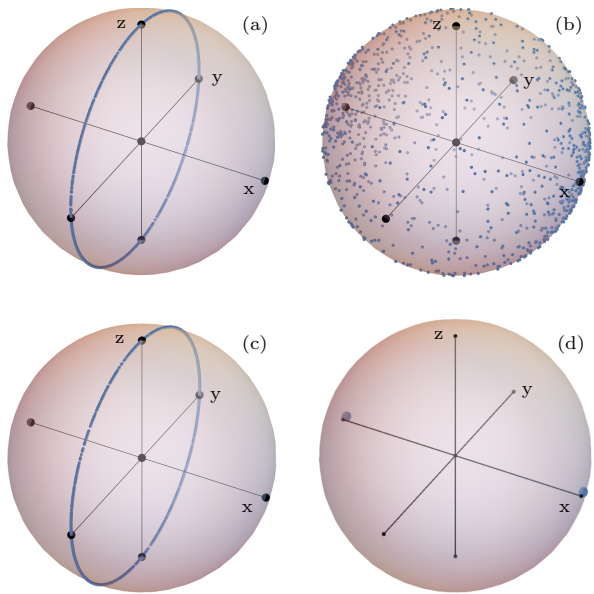
    \caption{{Endpoint ensembles}. Measurements {in $\vx$-} (a) and
      $\vz$-direction (c) confine the trajectory endpoints to a great
      circle around the $x$-axis. The $\vy$-ensemble (b) is spread
      over the whole Bloch sphere with a higher density around the
      $x$-eigenstates. The dichotomic ensemble shown in (d) was
      obtained {from} an adaptive scenario. We have plotted $10^3$
      points for each ensemble. The parameters are given in
      Eq.~(\ref{eq:parameters}).}\label{fig:ensembles}
  \end{figure}

{In the simplest measurement scenario Alice measures the same sharp
  observable $\So_\vn^\pm=\frac{1}{2}(\id\pm\vn\cdot\vsigma)$ on each
  subenvironment~\cite{brun,busch2013quantum}.}  {That is, each
  subenvironment is projected in the same basis after its collision.}
{In Fig.~\ref{fig:ensembles} we plot the endpoint ensembles on the
  Bloch sphere for measurement scenarios where Alice does the spin
  measurement {in} $\vx$-, $\vy$- and $\vz$-direction, respectively.}
The parameters used for the simulations are
\begin{align}
  \label{eq:parameters}
  {\gamma=1,\, \omega = 10,\,\delta t = 0.001,}
\end{align} {with $g=\sqrt{\gamma\delta t}$ and $f=\omega\delta t$.}
{Every point is the endpoint of a single discrete trajectory of length
  $N = 10^6$. We plot $10^3$ points.} {Here the system is
  initially prepared in the steady state of its reduced dynamics but
  note that in the limit of long trajectories ($l \rightarrow \infty$)
  the ensembles do not depend on the initial state.}

{Different choices for the measurement confine the endpoint ensembles
  to different regions on the surface of the Bloch sphere.} {While the
  $\vx$- and the $\vz$-ensemble are located on the great circle around
  the $x$-axis, the $\vy$-ensemble looks different.} The trajectory
endpoints are spread over the whole Bloch sphere with a higher density
close to the poles of the $x$-axis (see Fig.~\ref{fig:ensembles}
(a-c)).

  Since measurements in $\vx$- and $\vz$-direction lead to the same
  ensembles, these scenarios {will not violate the steering
    inequality} (\ref{eq:inequality}). {In contrast, if one of them is
    combined with the $\vy$-ensemble, the steering inequality
    (\ref{eq:inequality}) is violated.} {We have
    $\E_{\vx}[\langle \sigma_{\vy} \rangle^2] +\E_{\vx}[\langle
    \sigma_\vz \rangle^2] = 1$ because all points produced by
    measurements in $\vx$-direction are located on the great circle
    around the $x$-axis. For measurements in $\vy$-direction we obtain
    numerically
    $\E_{\vy}[\langle \sigma_\vx \rangle^2] = 0.546\pm 0.002$. Thus,
    there is a violation.}

  \subsection{Local adaptive scenario}\label{sec:local-adapt-scen}

  {The $\vx$-, $\vy$- and $\vz$-ensembles are not ideal with respect
    to the steering inequality (\ref{eq:inequality})}. A larger
  violation can be reached if the great-circle-ensemble (produced by
  measurements in $\vx$- or $\vz$-direction) is combined with an
  ensemble which only consist of points close to the $x$-eigenstates.

  {If the measurement direction for the subenvironments is not fixed,
    then it is possible to produce an ensemble which is close to such
    a dichotomic one~\cite{wiseman_quantum_1999}.}  {The driving
    rotates the system state around the $x$-axis.} Thus, we should
  look for a dichotomic ensemble whose states are relatively
  unaffected by the driving, that is, which are close to the $x$-axis.
  {The dichotomic pure state ensemble which satisfies the average
    reduced steady {state $\vr_{SS}$} and whose points are as close as
    possible to the $x$-axis consists of states}
  \begin{equation}
    \vr^{\pm} = \begin{pmatrix}
      \pm\sqrt{1-y_{SS}^2 - z_{SS}^2} \\ 
      y_{SS} \\ 
      z_{SS}
    \end{pmatrix},
  \end{equation}
  where $y_{SS}$ and $z_{SS}$ are taken from the steady state Bloch
  vector in Eq.~(\ref{eq:steady-bloch}).

  {To have a dichotomic endpoint {ensemble, necessarily, } as soon as
    the state of the trajectory is in the ensemble, it can either stay
    unaffected by the subsequent collision and measurement or jump to
    the other state in the ensemble.}  {This leads to the following
    conditions
    \begin{align}
      \label{eq:measurement-plus}
      \frac{\Tr_{\Anc_i}\left[(\id\otimes\So_{\vn_1}^{\pm})(\Q_i(\rho_\Sys^+\otimes\rho_{\Anc_i})\Q_i^\dagger)
      (\id\otimes\So_{\vn_1}^{\pm})\right]}{p_{\vn_1}^\pm}=&\rho_\Sys^\mp,\\
      \label{eq:measuremnt-minus}
      \frac{\Tr_{\Anc_i}\left[(\id\otimes\So_{\vn_2}^{\pm})(\Q_i(\rho_\Sys^-\otimes\rho_{\Anc_i})\Q_i^\dagger)
      (\id\otimes\So_{\vn_2}^{\pm})\right]}{p_{\vn_2}^\pm}=&\rho_\Sys^\pm,
    \end{align}
    where $\rho_\Sys^\pm=\frac{1}{2}(\id+\vr^\pm\cdot\vsigma)$,
    $p_{\vn_i}^\pm$ are measurement outcome probabilities and
    $\So_{\vn_1}$ and $\So_{\vn_2}$ are unknown observables to be
    found.}

  {The Eqs.~(\ref{eq:measurement-plus}),(\ref{eq:measuremnt-minus})
    are satisfied if the directions of the two spin observables are
    \begin{align}
      \vn_1 = \begin{pmatrix}
        0\\
        \sin(g)\\
        \cos(g)
      \end{pmatrix}, && \vn_2 = \begin{pmatrix}
        0\\
        -\sin(g)\\
        \cos(g)
      \end{pmatrix}.
    \end{align}}

  {The dichotomic ensemble can be produced from any initial state
    using the following measurement scenario.
    \begin{enumerate}
    \item Alice chooses first to measure the spin in one of the
      directions $\vn_i$, arbitrarily.
    \item As long as she obtains outcome $-1$ she keeps measuring to
      the chosen direction.\label{item:1}
    \item When she obtains an outcome $+1$ she will change to the
      other direction and continues at step \ref{item:1}.
    \end{enumerate}}

  {As shown in Fig.~\ref{fig:ensembles}\,(d), this measurement
    scenario produces the dichotomic ensemble.}  {Note that, as
    $\delta t\to 0$, our measurement scenario reduces to the adaptive
    quantum jump scheme of Ref. \cite{wiseman_quantum_1999}.}

  \subsection{Nonlocal scenario}
  \label{sec:nonlocal}
  We turn now to nonlocal scenarios as introduced in Section
  \ref{sec:nonlocal-scenarios}.  Let us assume that the system qubit
  is initially in a pure state (if it is in a mixed state, then we can
  apply a local measurement scenario until the trajectory has reached
  the surface of the Bloch ball and, hence, a pure state.). {Since all
    subenvironments are initially in the ground state and all
    interactions are unitary, the global joint state will remain pure
    for any number of collisions}.  {In the strong driving limit
    $\omega \gg \gamma$, the reduced state of the open system is
    almost maximally mixed $\rho_{SS} \approx \frac{1}{2}\mathbb{1}$.
    Therefore, in principle, it is possible to steer the system to any
    dichotomic pure state ensemble if one applies the right global
    measurement to the whole environment.} As already mentioned in
  Section \ref{sec:nonlocal-scenarios}, this is not in the spirit of a
  collision model. {Alternatively, we would like to collect the
    entanglement between the system and the whole environment in a
    single maximally entangled two-qubit state which would provide the
    same steering capabilities as the global joint pure state.}

  {To reach this aim, we add a two dimensional ancilla $\Cont$ which
    acts as a control system.}  {After each collision {between $\Sys$}
    and the subenvironment $\Anc_i$, the control system $\Cont$
    interacts with $\Anc_i$ with a unitary gate $\T_i$, after which
    $\Anc_i$ can be traced out (see Fig.~\ref{fig:cm-control})}.  {{We
      want} to find a universal $\T$-gate
    $\T_{\Cont\Anc_i}=\T_{\Cont\Anc}$}, such that
  \begin{equation}
    \rho_{\Sys\Cont}^\star = \Tr_{\Anc_i}[\T_{\Cont\Anc_i}\Q_{\Sys\Anc_i}(\rho_{\Sys\Cont}^\star \otimes \rho_{\Anc_i})\Q_{\Sys\Anc_i}^{\dagger} \T_{\Cont\Anc_i}^{\dagger}],
    \label{eq:bipartite-map}
  \end{equation}
  where $\rho_{\Sys\Cont}^{\star}$ is (i) a pure steady state for
  $\Sys+\Cont$ and therefore (ii) having the maximal amount of
  entanglement, with respect to the mixedness of $\rho_{SS}$. The
  subscripts of the unitaries {refer} to which subsystems they act.

  {To reduce the degrees of freedom in Eq.~(\ref{eq:bipartite-map}),
    we substitute an ansatz of the form
    $\Theta_{\Sys\Cont}^\star=\matt{1 & \vc^T\\ \vr_{SS} & R}$ in the
    Bloch representation for $\rho_{\Sys\Cont}^\star$~\cite{gamel}.}
  {$\Theta_{\Sys\Cont}^\star$ gives the correct steady state
    {$\vr_{SS}$} for the open system. $\vc$ and $R$ are as
    \begin{align*}
      \vc = \begin{pmatrix} 0 \\ 0 \\ \cos \alpha
      \end{pmatrix},\, R = \begin{pmatrix}
        0 & \sin\alpha & 0 \\
        -\sin\alpha\cos\beta & 0 & -\sin\beta\\
        \sin\alpha\sin\beta & 0 & -\cos\beta
      \end{pmatrix},
    \end{align*}
    where $\alpha = \arccos(|\vr_{SS}|)$ and
    $\beta = \arctan\left(\frac{y_{SS}}{z_{SS}}\right) +
    \pi$~\footnote{{The suitable ansatz was found by considering the
        given constraints and numerical estimations. This ansatz is
        not unique but it is chosen such that it leads to a continuous
        limit.}}}.  {The above choice for $\vc$ and $R$ satisfies the
    condition (i) and therefore also (ii).}

  {In order to find $\T_{\Cont\Anc}$, {we evolve} the pure state
    $\rho_{\Sys\Cont}^\star$ =
    $\ket{\psi_{\Sys\Cont}^\star}\bra{\psi_{\Sys\Cont}^\star}$ over a
    single collision}
  \begin{equation}
    \ket{\psi_ {\Sys\Cont\Anc}'} = \Q(\ket{\psi_ {\Sys\Cont}^\star} \otimes \ket{0}).
  \end{equation}
  In order to keep $\Sys+\Cont$ maximally entangled we have to find a
  $\T_{\Cont\Anc}$ such that the subenvironment is decoupled from the
  system and the control, that is
  \begin{equation}
    (\id_\Sys \otimes \T_{\Cont\Anc})\ket{\psi_ {\Sys\Cont\Anc}'} 
    = \ket{\psi_ {\Sys\Cont}'} \otimes \ket{0_\Anc}.
    \label{eq:singular-decomposition}
  \end{equation}
  $\T_{\Cont\Anc}$ can be constructed by writing
  $\ket{\psi_ {\Sys\Cont\Anc}'}$ in its Schmidt decomposition
  \begin{equation}
    \ket{\psi_{\Sys\Cont\Anc}'} = 
    \lambda_1  \ket{\psi_\Sys^1} \otimes \ket{\psi_{\Cont\Anc}^1} 
    + \lambda_2 \ket{\psi_\Sys^2} \otimes \ket{\psi_{\Cont\Anc}^2}.
  \end{equation} 
  The vectors $\ket{\psi_{\Cont\Anc}^1}$ and
  $\ket{\psi_{\Cont\Anc}^2}$ are orthogonal. Together with two further
  orthogonal vectors $\ket{\psi_{\Cont\Anc}^3}$ and
  $\ket{\psi_{\Cont\Anc}^4}$ they form a basis of the two-qubit
  Hilbert space $\mathcal{H}_{\Cont\Anc}$. Thus, we can construct the
  following unitary operator $\T_{\Cont\Anc}$ which decouples the
  subenvironment:
  \begin{align}
    \T_{\Cont\Anc} = &\ket{0_\Cont 0_\Anc}\bra{\psi_{\Cont\Anc}^1}
                       + \ket{1_C 0_A}\bra{\psi_{\Cont\Anc}^2}\\
                     &+\ket{0_\Cont 1_\Anc}\bra{\psi_{\Cont\Anc}^3} + \ket{1_\Cont
                       1_\Anc}\bra{\psi_{\Cont\Anc}^4}.
  \end{align}
  We find
  \begin{equation}
    \label{eq:T-gate}
    \T_{\Cont\Anc} = \T_2\T_1 =e^{-i \omega \delta t \, S_2} \, e^{-i \sqrt{\gamma \delta t} \, S_1},
  \end{equation}
  where $S_1$ and $S_2$ are given in the Appendix
  \ref{sec:universal-t-gate}.  This unitary $\T_{\Cont\Anc}$ together
  with the $\rho_{\Sys\Cont}^\star$ satisfies
  Eq.~(\ref{eq:bipartite-map}).

  If Alice uses this nonlocal measurement scenario, she eventually
  holds one part of a maximally entangled state. As soon as Bob stops
  the experiment, Alice can steer Bob's system to an arbitrary
  dichotomic ensemble by performing a projective measurement on the
  control qubit.

  {It is possible to obtain a continuous limit for the collision model
    for $\Sys+\Cont$} which leads to a GKSL master equation for the
  two-qubit state $\rho_{\Sys\Cont}$.  {Proceeding as in
    Sec.~\ref{sec:example}, the Hamiltonian and the Lindblad operator
    turn out to be
    \begin{equation}
      \begin{split}
        H_{\Sys\Cont} = &-\omega(\mathbb{1} \otimes \sigmay) + \omega R_{33} (\sigmax \otimes \sigmaz) + \omega (\sigmax \otimes \mathbb{1}) \\
        &+\frac{1}{4}R_{33} \gamma (\sigmax \otimes \sigmax) +
        \frac{1}{4} \gamma (\sigmay \otimes \sigmay),
      \end{split}
    \end{equation}
    \begin{equation}
      \begin{split}
        L_{\Sys\Cont} = &-\frac{2 R_{33} \omega}{\sqrt{\gamma}}
        (\mathbb{1} \otimes \sigmaz) - \frac{1}{2} R_{33}
        \sqrt{\gamma} (\mathbb{1} \otimes \sigmax) \\
        &+\frac{i}{2}\sqrt{\gamma}(\mathbb{1} \otimes \sigmay) -
        i\sqrt{\gamma} (\sigmam \otimes \mathbb{1}),
      \end{split}
    \end{equation}}
  with $R_{33} =
  -\frac{\gamma }{\sqrt{\gamma ^2+16 \omega ^2}}$.

  {The GKSL master equation for the open system only,
    Eq.~(\ref{eq:gksl}), is embedded into this {two-qubit} GKSL master
    equation.}  {Its steady state solution is (in Bloch
    representation):}
  \begin{equation}
    \Theta_{SS} = \begin{pmatrix}
      1 & 0 & 0 & \frac{\sqrt{16 c^2+1}}{8 c^2+1} \\
      0 & 0 & \frac{8 c^2}{8 c^2+1} & 0 \\
      \frac{4 c}{8 c^2+1} & -\frac{8 c^2}{\left(8 c^2+1\right) \sqrt{16 c^2+1}} & 0 & \frac{4 c}{\sqrt{16 c^2+1}} \\
      \frac{1}{-8 c^2-1} & -\frac{32 c^3}{\left(8 c^2+1\right) \sqrt{16 c^2+1}} & 0 & -\frac{1}{\sqrt{16 c^2+1}} \\
    \end{pmatrix},
  \end{equation}
  where $c=\omega/\gamma$. Thus, in the continuous limit, this state
  only depends on the ratio between the driving and the damping.

  {In the collision model the creation of the entanglement of the
    bipartite state $\rho_{\Sys\Cont}$ is mediated by the
    subenvironments and the control and the system never interact
    directly. Interestingly, in the continuous limit the Hamiltonian
    $H_{\Sys\Cont}$ is of interaction type.}

\section{Coupling to a Thermal Bath}\label{sec:thermal}
{We consider now the general case
where the environment is allowed to be initially
in a mixed state. The question whether the system
is entangled with the environment cannot be answered by the simple
argument which holds for pure states only.} 
In the collision model we replace the pure initial state of the
subenvironments with a thermal one
$\rho_{\Anc} =\bigotimes_i \frac{1}{2}(\mathbb{1}^i+\eta \sigmaz^i)$,
where $\eta \in (-1,0)$ is a temperature parameter{, related to the
  Boltzmann factor~\cite{scarani-thermalizing}.}  {The collision model
  now describes a damped driven qubit coupled to a thermal
  bath. Correspondingly, the time-continuous limit of our model gives the GKSL master equation
  \begin{equation}
    \dot{\rho}_\Sys = -i\omega[\sigmax,\rho_\Sys] 
    + \gamma \frac{1-\eta}{2}\mathcal{D}[\sigmam]\rho_\Sys 
    + \gamma \frac{1+\eta}{2}\mathcal{D}[\sigmap]\rho_\Sys,
    \label{eq:gksl-thermal}
  \end{equation}}
where
$\mathcal{D}[L]\rho = L\rho L^{\dagger} -
\frac{1}{2}\{\rho,L^{\dagger}L\}$.

\begin{figure}[ht]
    \def\svgwidth{.9\columnwidth}
    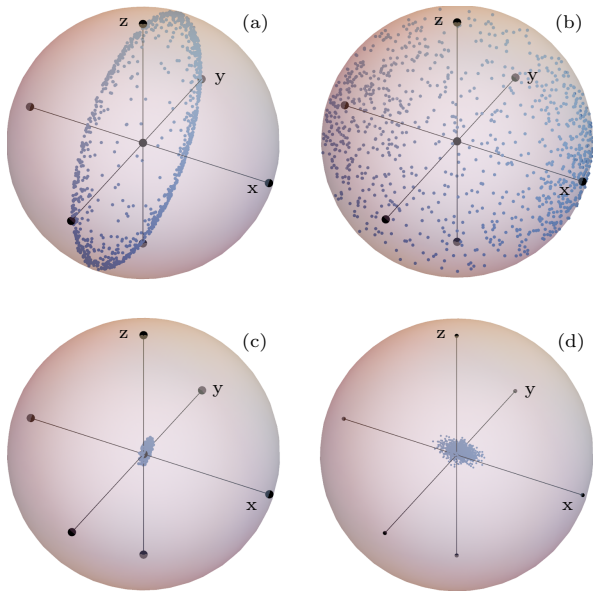
    \caption{If the subenvironments are in a thermal state, the
      endpoint ensembles are not pure anymore. For measurements in
      $\vx$- and $\vy$-direction (a,b) the ensembles resemble the
      ensembles obtained in the vacuum case. By contrast, measurements
      in $\vz$-direction (c) as well as the adaptive scenario (d) lead
      to ensembles which are concentrated close to the center of the
      Bloch sphere, that is the fully mixed state. The ensembles (c)
      and (d) cannot demonstrate steering but (a) and (b) violate a
      steering inequality if the temperature is not too
      high.}\label{fig:thermal-ensembles}
  \end{figure}

{The ensembles produced by any measurement scenario in a collision
  model with thermal subenvironments will certainly not be pure.} On
the other hand, the steering inequality (\ref{eq:inequality}) does not
rely on pure ensembles but only on the ensemble averages for
$\langle \operatorname{\sigma_\vn} \rangle^2$.  In
Fig.~\ref{fig:thermal-ensembles} we show endpoint ensembles for
measurements in $\vx$-, $\vy$- and $\vz$-direction, respectively, and
for the adaptive scenario for a temperature $\eta = -0.9$. The
parameters are chosen again as in Eq.~(\ref{eq:parameters}).

  We see that some measurement scenarios are more robust against
  thermal noise than others. In all cases {the thermal
    subenvironments} lead to mixed conditional states in Bob's system,
  since Alice does not fully know the state of her system before the
  collisions.  {While the purity of the conditional states in the
    $\vx$- or $\vy$-ensemble decreases slowly with increasing
    temperature, the conditional states in the $\vz$- and in the
    adaptive ensemble lose their purity even if the temperature is
    very low.}  Accordingly, the $\vx$- and $\vy$-ensembles are most
  suitable in order to demonstrate steering. We define the
  \textit{steerability} $\Delta S$:
  \begin{equation}
    \Delta S = 
    \E_\vx[\langle \sigma_\vy \rangle^2] 
    + \E_\vx[\langle \sigma_\vz \rangle^2] 
    +  \E_\vy[\langle \sigma_\vx \rangle^2] - 1.
  \end{equation}


  {Steering is successfully demonstrated whenever $\Delta S > 0$. The
    steerability $\Delta S$ decreases with increasing
    temperature. This is not surprising because the thermal
    subenvironments induce noise. Numerical simulations show that
    there is a critical value
    {$\eta_{\textup{crit}} = -0.72\pm0.01$} up to which the system
    can be steered by the chosen measurement scenarios. This value is
    an upper bound for a wide range of the parameters $\delta t$,
    $\omega$ and $\gamma$ in the discrete model. In particular, this
    $\eta_{\textup{crit}}$ is also reached for strong driving
    $\omega \gg \gamma$ and small time steps $\omega \delta t \ll 1$,
    where the discrete collision model is a good numerical
    approximation of the related continuous process. However, we
    recall that $\eta_{\textup{crit}}$ depends on the chosen
    measurement scenarios and therefore must not be seen as a
    universal bound for steerability of the system. There might be
    better measurement scenarios and inequalities which are able to
    detect steering at even higher temperatures.  }



  {Steerability implies entanglement, therefore, at least up to this
    critical temperature the system and the environment build up
    entanglement during the interaction. However, we would like to
    emphasize that there is a big difference between the pure state
    case (coupling to a vacuum bath) and the thermal case presented in
    this section. The single pure subenvironments entangle with the
    system during their collision, whereas the thermal subenvironments
    do not. More precisely, the two-qubit entanglement between the
    system and a single subenvironment right after their interaction
    vanishes in general if the subenvironment is in a thermal state
    but is nonzero if the subenvironment is initially in a pure vacuum
    state. Let us assume that the system starts in its steady state
    $\rho_{SS}$ and, thus, the reduced state does not change during
    the collisions. Then the bipartite state of the system and a
    certain subenvironment $\Anc_i$ right after their collision is the
    same for all $i$. Collisions happening after the interaction
    between $\Sys$ and $\Anc_i$ cannot increase the entanglement
    between these two qubits. Therefore, we only have to check whether
    the bipartite state
    $\rho_{\Sys\Anc_i}^{SS} =
    \Q_i(\rho_{SS}\otimes\rho_{\Anc_i})\Q_i^\dagger$ is entangled. The
    steady state $\rho_{SS}$ depends on the temperature parameter
    $\eta$ and the time step $\delta t$ and so does the bipartite
    entanglement~\footnote{{We set $\gamma=1$ since it defines only a
        scaling of $\delta t$ in the discrete model. Then the steady
        state also depends on the driving parameter $\omega$. However,
        its influence on the bipartite entanglement is negligible over
        a wide parameter range. All numerical results presented here
        are calculated for a fixed driving $\omega=10$.}}. The Bloch
    vector corresponding to $\rho_{SS}$ can be found in the
    appendix. In Fig.~\ref{fig:bipartite-entanglement} we show the
    two-qubit entanglement between the system and single
    subenvironments for different temperatures and time steps.  }

\begin{figure}[ht]
  \def\svgwidth{.9\columnwidth}
  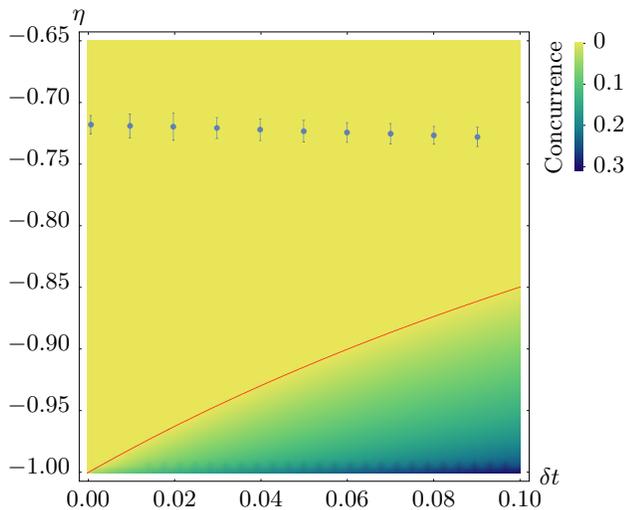
  \caption{{We plot the concurrence between the system $\Sys$ and a
      single subenvironment $\Anc_i$ after their collision. $\Sys$ is
      initially prepared in the steady state $\rho_{SS}$ and $\Anc_i$
      in the thermal state
      $\rho_{\Anc_i}=\frac{1}{2} (\id+\eta\sigmaz)$. The bipartite
      entanglement depends on the temperature parameter $\eta$ and the
      time step $\delta t$. In the parameter region beneath the thin
      line the system becomes entangled with single subenvironments,
      in the region above it does not. In particular the critical
      temperatures for steering $\eta_{\textup{crit}}$ (blue dots) are
      in the parameter region without bipartite
      entanglement.}}\label{fig:bipartite-entanglement}
\end{figure}

{For finite time steps $\delta t$ the system always builds up
  two-qubit entanglement with vacuum subenvironments ($\eta=-1$). If
  the bath is thermal, then the temperature up to which bipartite
  entanglement occurs depends on the time step. The smaller the time
  step the smaller the temperature $\eta$ above which the two-qubit
  entanglement vanishes. In particular the critical temperature
  $\eta_{\textup{crit}}$ for the steering task is much higher then the
  temperature which allows bipartite entanglement. The
possibility of steering in the thermal case therefore implies
that the environment as a whole entangles to the system.
{It would be surprising if a nonlocal measurement scenario, as seen in
  Sec.~\ref{sec:nonlocal}, which is based on accumulating two-qubit
  entanglement, could be constructed for the thermal case.}

\section{Conclusion}\label{sec:conclusion}

We have shown how the concept of quantum steering of an open quantum
system can be discussed elegantly in collision models. The approach
provides transparent insights for the choice of actual measurements
for the steering tasks. We have argued that local non-adaptive and
nonlocal measurement scenarios are instances of bipartite steering,
whereas a local adaptive measurement scenario can be seen as a
multipartite steering task.

With the help of a concrete example --- a coherently driven qubit
damped by a bath --- we have shown how different measurement scenarios
affect the system and how possible endpoint ensembles look like. We
have introduced a special form of nonlocal measurements on the
environment which fits naturally in the framework of a collision
model.  In the case of a vacuum bath, such a scenario can be used to
accumulate entanglement between the system and the control state.  In
the time-continuous limit this discrete model leads to a Markovian
master equation for two qubits, embedding the reduced system dynamics.

In Sec.~\ref{sec:thermal} we have shown that steering is not
restricted to the pure state case only, but can also be demonstrated
if the environment is in a thermal, that is a mixed, initial
state. The robustness against this thermal noise strongly depends on
the used measurement scenarios and steering becomes harder with
increasing temperature.  For two simple scenarios we could numerically
estimate a critical value up to which steering can be witnessed. In
this temperature range the system and its environment evolve towards
an entangled state. Interestingly, there is no two-qubit entanglement
between the system and single subenvironments in the thermal
case. This leads to the conclusion that steering is based on entanglement between the
open system and the whole bath. For deeper insights
in how the system gets entangled with its environment it would be
necessary to investigate $n$-qubit entanglement in the global joint
state which is in general quite involved, both analytically and
numerically.

\begin{acknowledgments}
  The authors would like to thank Ana Costa, Chau Nguyen and Roope
  Uola for many fruitful discussions. We are also grateful to Howard Wiseman for
  helpful comments on the manuscript.
\end{acknowledgments}

\appendix
\section{Universal T-gate}\label{sec:universal-t-gate}
The exact expressions for the matrices $S_1$ and $S_2$ in the
decomposition for the $\T$-gate in Eq.~(\ref{eq:T-gate}) expressed in
the computational basis $\{\ket{11},\ket{10},\ket{01},\ket{00}\}$ are
\begin{widetext}
  \begin{align}
    S_1 &= \begin{pmatrix}
      0 & -\frac{2 i R_{33} \omega }{\gamma } & 0 & -\frac{1}{2} i (R_{33}-1) \\
      \frac{2 i R_{33} \omega }{\gamma } & 0 & \frac{1}{2} i (R_{33}+1) & 0 \\
      0 & -\frac{1}{2} i (R_{33}+1) & 0 & \frac{2 i R_{33} \omega }{\gamma } \\
      \frac{1}{2} i (R_{33}-1) & 0 & -\frac{2 i R_{33} \omega }{\gamma } & 0 \\
    \end{pmatrix}, &&\begin{array}{c}
                       \phantom{\,} \\
                       \phantom{\,} \\
                       \phantom{\,} \\
                       \phantom{\,} \\
                       \phantom{\,} \\
                       \phantom{\,}
                     \end{array}  \\
    S_2 &= \begin{pmatrix}
      0 & 0 & i R_{33} & 0 \\
      0 & 0 & 0 & i \\
      -i R_{33} & 0 & 0 & 0 \\
      0 & -i & 0 & 0 \\
    \end{pmatrix}, \quad\quad\quad\quad\quad R_{33} =
    -\frac{\gamma}{\sqrt{\gamma^2+16 \omega^2}}.
  \end{align}
\end{widetext}

\section{Steady state Bloch vector for a thermal environment}

\begin{equation}
  \vr_{SS} = \begin{pmatrix}
    0 \\
    -\frac{4 \eta \sin^2 \left(\frac{\sqrt{\gamma \delta t}}{2}\right)
      \cos \left(\sqrt{\gamma  \delta t}\right) \sin (2 \delta t \omega
      )}{-4 \cos \left(\sqrt{\gamma \delta t}\right) \cos^2(\delta t
      \omega )+\cos \left(2 \sqrt{\gamma \delta t}\right)+3} \\
    -\frac{4 \eta \sin^2\left(\frac{\sqrt{\gamma \delta t}}{2}\right)
      \left(\cos \left(\sqrt{\gamma  \delta t}\right)
        \cos (2 \delta t \omega )-1\right)}{-4 \cos \left(\sqrt{\gamma  \delta t}\right)
      \cos^2(\delta t \omega )+\cos \left(2 \sqrt{\gamma  \delta t}\right)+3} \\
  \end{pmatrix}.
\end{equation}
\vspace{1pt}

\bibliography{bibliography}
\end{document}

%% file: images/pdf/cm-simple.pdf_tex
\begingroup%
  \makeatletter%
  \providecommand\color[2][]{%
    \errmessage{(Inkscape) Color is used for the text in Inkscape, but the package 'color.sty' is not loaded}%
    \renewcommand\color[2][]{}%
  }%
  \providecommand\transparent[1]{%
    \errmessage{(Inkscape) Transparency is used (non-zero) for the text in Inkscape, but the package 'transparent.sty' is not loaded}%
    \renewcommand\transparent[1]{}%
  }%
  \providecommand\rotatebox[2]{#2}%
  \ifx\svgwidth\undefined%
    \setlength{\unitlength}{483.63330078bp}%
    \ifx\svgscale\undefined%
      \relax%
    \else%
      \setlength{\unitlength}{\unitlength * \real{\svgscale}}%
    \fi%
  \else%
    \setlength{\unitlength}{\svgwidth}%
  \fi%
  \global\let\svgwidth\undefined%
  \global\let\svgscale\undefined%
  \makeatother%
  \begin{picture}(1,0.25004382)%
    \put(0,0){\includegraphics[width=\unitlength]{cm-simple.pdf}}%
    \put(0.3182488,0.1798913){\color[rgb]{0,0,0}\makebox(0,0)[b]{\smash{$\Q_1$}}}%
    \put(0.12149705,0.22395566){\color[rgb]{0,0,0}\makebox(0,0)[rb]{\smash{$\rho_{\Sys_{\phantom{1}}}$}}}%
    \put(0.12422641,0.08216221){\color[rgb]{0,0,0}\makebox(0,0)[rb]{\smash{$\rho_{\Anc_2}$}}}%
    \put(0.12422641,0.01117244){\color[rgb]{0,0,0}\makebox(0,0)[rb]{\smash{$\rho_{\Anc_3}$}}}%
    \put(0.12422641,0.15315198){\color[rgb]{0,0,0}\makebox(0,0)[rb]{\smash{$\rho_{\Anc_1}$}}}%
    \put(0.87074233,0.22397731){\color[rgb]{0,0,0}\makebox(0,0)[lb]{\smash{$\rho_\Sys'$}}}%
    \put(0.47704678,0.14750156){\color[rgb]{0,0,0}\makebox(0,0)[b]{\smash{$\Q_2$}}}%
    \put(0.63584471,0.11637209){\color[rgb]{0,0,0}\makebox(0,0)[b]{\smash{$\Q_3$}}}%
  \end{picture}%
\endgroup%

%% file: images/pdf/cm-measurement.pdf_tex
\begingroup%
  \makeatletter%
  \providecommand\color[2][]{%
    \errmessage{(Inkscape) Color is used for the text in Inkscape, but the package 'color.sty' is not loaded}%
    \renewcommand\color[2][]{}%
  }%
  \providecommand\transparent[1]{%
    \errmessage{(Inkscape) Transparency is used (non-zero) for the text in Inkscape, but the package 'transparent.sty' is not loaded}%
    \renewcommand\transparent[1]{}%
  }%
  \providecommand\rotatebox[2]{#2}%
  \ifx\svgwidth\undefined%
    \setlength{\unitlength}{483.63330078bp}%
    \ifx\svgscale\undefined%
      \relax%
    \else%
      \setlength{\unitlength}{\unitlength * \real{\svgscale}}%
    \fi%
  \else%
    \setlength{\unitlength}{\svgwidth}%
  \fi%
  \global\let\svgwidth\undefined%
  \global\let\svgscale\undefined%
  \makeatother%
  \begin{picture}(1,0.25004382)%
    \put(0,0){\includegraphics[width=\unitlength]{cm-measurement.pdf}}%
    \put(0.3182488,0.1854976){\color[rgb]{0,0,0}\makebox(0,0)[b]{\smash{$\Q_1$}}}%
    \put(0.12149705,0.22395566){\color[rgb]{0,0,0}\makebox(0,0)[rb]{\smash{$\rho_{\Sys_{\phantom{i}}}$}}}%
    \put(0.12422641,0.08216221){\color[rgb]{0,0,0}\makebox(0,0)[rb]{\smash{$\rho_{\Anc_2}$}}}%
    \put(0.12422641,0.01117244){\color[rgb]{0,0,0}\makebox(0,0)[rb]{\smash{$\rho_{\Anc_3}$}}}%
    \put(0.12422641,0.15315198){\color[rgb]{0,0,0}\makebox(0,0)[rb]{\smash{$\rho_{\Anc_1}$}}}%
    \put(0.87074233,0.22397731){\color[rgb]{0,0,0}\makebox(0,0)[lb]{\smash{$\rho_\Sys'$}}}%
    \put(0.47704678,0.1510304){\color[rgb]{0,0,0}\makebox(0,0)[b]{\smash{$\Q_2$}}}%
    \put(0.63584471,0.11527226){\color[rgb]{0,0,0}\makebox(0,0)[b]{\smash{$\Q_3$}}}%
  \end{picture}%
\endgroup%

%% file: images/pdf/cm-control.pdf_tex
\begingroup%
  \makeatletter%
  \providecommand\color[2][]{%
    \errmessage{(Inkscape) Color is used for the text in Inkscape, but the package 'color.sty' is not loaded}%
    \renewcommand\color[2][]{}%
  }%
  \providecommand\transparent[1]{%
    \errmessage{(Inkscape) Transparency is used (non-zero) for the text in Inkscape, but the package 'transparent.sty' is not loaded}%
    \renewcommand\transparent[1]{}%
  }%
  \providecommand\rotatebox[2]{#2}%
  \ifx\svgwidth\undefined%
    \setlength{\unitlength}{483.63330078bp}%
    \ifx\svgscale\undefined%
      \relax%
    \else%
      \setlength{\unitlength}{\unitlength * \real{\svgscale}}%
    \fi%
  \else%
    \setlength{\unitlength}{\svgwidth}%
  \fi%
  \global\let\svgwidth\undefined%
  \global\let\svgscale\undefined%
  \makeatother%
  \begin{picture}(1,0.31949861)%
    \put(0,0){\includegraphics[width=\unitlength]{cm-control.pdf}}%
    \put(0.31762034,0.22167641){\color[rgb]{0,0,0}\makebox(0,0)[b]{\smash{$\Q_1$}}}%
    \put(0.12149707,0.29472589){\color[rgb]{0,0,0}\makebox(0,0)[rb]{\smash{$\rho_{\Sys_{\phantom{1}}}$}}}%
    \put(0.12149707,0.15293244){\color[rgb]{0,0,0}\makebox(0,0)[rb]{\smash{$\rho_{\Anc_1}$}}}%
    \put(0.12149707,0.08194267){\color[rgb]{0,0,0}\makebox(0,0)[rb]{\smash{$\rho_{\Anc_2}$}}}%
    \put(0.12149707,0.22392221){\color[rgb]{0,0,0}\makebox(0,0)[rb]{\smash{$\rho_{\Cont_{\phantom{1}}}$}}}%
    \put(0.87074233,0.29474754){\color[rgb]{0,0,0}\makebox(0,0)[lb]{\smash{$\rho_\Sys'$}}}%
    \put(0.47651387,0.18638797){\color[rgb]{0,0,0}\makebox(0,0)[b]{\smash{$\Q_2$}}}%
    \put(0.63539208,0.15109953){\color[rgb]{0,0,0}\makebox(0,0)[b]{\smash{$\Q_3$}}}%
    \put(0.39660819,0.18569096){\color[rgb]{0,0,0}\makebox(0,0)[b]{\smash{$\T_1$}}}%
    \put(0.55375203,0.15040252){\color[rgb]{0,0,0}\makebox(0,0)[b]{\smash{$\T_2$}}}%
    \put(0.12149707,0.01136582){\color[rgb]{0,0,0}\makebox(0,0)[rb]{\smash{$\rho_{\Anc_3}$}}}%
    \put(0.71089591,0.11568684){\color[rgb]{0,0,0}\makebox(0,0)[b]{\smash{$\T_3$}}}%
  \end{picture}%
\endgroup%

%% file: images/pdf/ensembles.pdf_tex
\begingroup%
  \makeatletter%
  \providecommand\color[2][]{%
    \errmessage{(Inkscape) Color is used for the text in Inkscape, but the package 'color.sty' is not loaded}%
    \renewcommand\color[2][]{}%
  }%
  \providecommand\transparent[1]{%
    \errmessage{(Inkscape) Transparency is used (non-zero) for the text in Inkscape, but the package 'transparent.sty' is not loaded}%
    \renewcommand\transparent[1]{}%
  }%
  \providecommand\rotatebox[2]{#2}%
  \ifx\svgwidth\undefined%
    \setlength{\unitlength}{1500.00005767bp}%
    \ifx\svgscale\undefined%
      \relax%
    \else%
      \setlength{\unitlength}{\unitlength * \real{\svgscale}}%
    \fi%
  \else%
    \setlength{\unitlength}{\svgwidth}%
  \fi%
  \global\let\svgwidth\undefined%
  \global\let\svgscale\undefined%
  \makeatother%
  \begin{picture}(1,1)%
    \put(0,0){\includegraphics[width=\unitlength,page=1]{ensembles.pdf}}%
    \put(0.39995716,0.98619577){\color[rgb]{0,0,0}\makebox(0,0)[lt]{\begin{minipage}{0.11140359\unitlength}\raggedright \scriptsize{(a)}\end{minipage}}}%
    \put(1.90833051,0.7749303){\color[rgb]{0,0,0}\makebox(0,0)[lt]{\begin{minipage}{0.05978996\unitlength}\raggedright \end{minipage}}}%
    \put(0.93458889,0.98619577){\color[rgb]{0,0,0}\makebox(0,0)[lt]{\begin{minipage}{0.11140359\unitlength}\raggedright \scriptsize{(b)}\end{minipage}}}%
    \put(0.39995716,0.44123299){\color[rgb]{0,0,0}\makebox(0,0)[lt]{\begin{minipage}{0.11140359\unitlength}\raggedright \scriptsize{(c)}\end{minipage}}}%
    \put(0.93846303,0.44123299){\color[rgb]{0,0,0}\makebox(0,0)[lt]{\begin{minipage}{0.11140359\unitlength}\raggedright \scriptsize{(d)}\end{minipage}}}%
    \put(0.3459236,0.34452961){\color[rgb]{0,0,0}\makebox(0,0)[lt]{\begin{minipage}{0.11140359\unitlength}\raggedright \scriptsize{y}\end{minipage}}}%
    \put(0.40008945,0.15273466){\color[rgb]{0,0,0}\makebox(0,0)[lt]{\begin{minipage}{0.11140359\unitlength}\raggedright \scriptsize{x}\end{minipage}}}%
    \put(0.18322789,0.44122491){\color[rgb]{0,0,0}\makebox(0,0)[lt]{\begin{minipage}{0.11140359\unitlength}\raggedright \scriptsize{z}\end{minipage}}}%
    \put(0.87809292,0.35352967){\color[rgb]{0,0,0}\makebox(0,0)[lt]{\begin{minipage}{0.11140359\unitlength}\raggedright \scriptsize{y}\end{minipage}}}%
    \put(0.94225877,0.15273466){\color[rgb]{0,0,0}\makebox(0,0)[lt]{\begin{minipage}{0.11140359\unitlength}\raggedright \scriptsize{x}\end{minipage}}}%
    \put(0.72839722,0.44822495){\color[rgb]{0,0,0}\makebox(0,0)[lt]{\begin{minipage}{0.11140359\unitlength}\raggedright \scriptsize{z}\end{minipage}}}%
    \put(0.88109292,0.88264859){\color[rgb]{0,0,0}\makebox(0,0)[lt]{\begin{minipage}{0.11140359\unitlength}\raggedright \scriptsize{y}\end{minipage}}}%
    \put(0.94225877,0.69085364){\color[rgb]{0,0,0}\makebox(0,0)[lt]{\begin{minipage}{0.11140359\unitlength}\raggedright \scriptsize{x}\end{minipage}}}%
    \put(0.72539722,0.97734389){\color[rgb]{0,0,0}\makebox(0,0)[lt]{\begin{minipage}{0.11140359\unitlength}\raggedright \scriptsize{z}\end{minipage}}}%
    \put(0.3487701,0.88764859){\color[rgb]{0,0,0}\makebox(0,0)[lt]{\begin{minipage}{0.11140359\unitlength}\raggedright \scriptsize{y}\end{minipage}}}%
    \put(0.40293594,0.69585364){\color[rgb]{0,0,0}\makebox(0,0)[lt]{\begin{minipage}{0.11140359\unitlength}\raggedright \scriptsize{x}\end{minipage}}}%
    \put(0.18607439,0.98034389){\color[rgb]{0,0,0}\makebox(0,0)[lt]{\begin{minipage}{0.11140359\unitlength}\raggedright \scriptsize{z}\end{minipage}}}%
  \end{picture}%
\endgroup%

%% file: images/pdf/ensembles-thermal.pdf_tex
\begingroup%
  \makeatletter%
  \providecommand\color[2][]{%
    \errmessage{(Inkscape) Color is used for the text in Inkscape, but the package 'color.sty' is not loaded}%
    \renewcommand\color[2][]{}%
  }%
  \providecommand\transparent[1]{%
    \errmessage{(Inkscape) Transparency is used (non-zero) for the text in Inkscape, but the package 'transparent.sty' is not loaded}%
    \renewcommand\transparent[1]{}%
  }%
  \providecommand\rotatebox[2]{#2}%
  \ifx\svgwidth\undefined%
    \setlength{\unitlength}{1500.00005767bp}%
    \ifx\svgscale\undefined%
      \relax%
    \else%
      \setlength{\unitlength}{\unitlength * \real{\svgscale}}%
    \fi%
  \else%
    \setlength{\unitlength}{\svgwidth}%
  \fi%
  \global\let\svgwidth\undefined%
  \global\let\svgscale\undefined%
  \makeatother%
  \begin{picture}(1,1)%
    \put(0,0){\includegraphics[width=\unitlength,page=1]{ensembles-thermal.pdf}}%
    \put(0.39995716,0.98619577){\color[rgb]{0,0,0}\makebox(0,0)[lt]{\begin{minipage}{0.11140359\unitlength}\raggedright \scriptsize{(a)}\end{minipage}}}%
    \put(1.90833051,0.7749303){\color[rgb]{0,0,0}\makebox(0,0)[lt]{\begin{minipage}{0.05978996\unitlength}\raggedright \end{minipage}}}%
    \put(0.93458889,0.98619577){\color[rgb]{0,0,0}\makebox(0,0)[lt]{\begin{minipage}{0.11140359\unitlength}\raggedright \scriptsize{(b)}\end{minipage}}}%
    \put(0.39995716,0.44123299){\color[rgb]{0,0,0}\makebox(0,0)[lt]{\begin{minipage}{0.11140359\unitlength}\raggedright \scriptsize{(c)}\end{minipage}}}%
    \put(0.93846303,0.44123299){\color[rgb]{0,0,0}\makebox(0,0)[lt]{\begin{minipage}{0.11140359\unitlength}\raggedright \scriptsize{(d)}\end{minipage}}}%
    \put(0.34992363,0.35352967){\color[rgb]{0,0,0}\makebox(0,0)[lt]{\begin{minipage}{0.11140359\unitlength}\raggedright \scriptsize{y}\end{minipage}}}%
    \put(0.40708949,0.15273466){\color[rgb]{0,0,0}\makebox(0,0)[lt]{\begin{minipage}{0.11140359\unitlength}\raggedright \scriptsize{x}\end{minipage}}}%
    \put(0.19022794,0.44722494){\color[rgb]{0,0,0}\makebox(0,0)[lt]{\begin{minipage}{0.11140359\unitlength}\raggedright \scriptsize{z}\end{minipage}}}%
    \put(0.88309292,0.35152965){\color[rgb]{0,0,0}\makebox(0,0)[lt]{\begin{minipage}{0.11140359\unitlength}\raggedright \scriptsize{y}\end{minipage}}}%
    \put(0.94725877,0.15273466){\color[rgb]{0,0,0}\makebox(0,0)[lt]{\begin{minipage}{0.11140359\unitlength}\raggedright \scriptsize{x}\end{minipage}}}%
    \put(0.73239722,0.44722494){\color[rgb]{0,0,0}\makebox(0,0)[lt]{\begin{minipage}{0.11140359\unitlength}\raggedright \scriptsize{z}\end{minipage}}}%
    \put(0.88709293,0.88764859){\color[rgb]{0,0,0}\makebox(0,0)[lt]{\begin{minipage}{0.11140359\unitlength}\raggedright \scriptsize{y}\end{minipage}}}%
    \put(0.94225877,0.69085364){\color[rgb]{0,0,0}\makebox(0,0)[lt]{\begin{minipage}{0.11140359\unitlength}\raggedright \scriptsize{x}\end{minipage}}}%
    \put(0.72839722,0.98234389){\color[rgb]{0,0,0}\makebox(0,0)[lt]{\begin{minipage}{0.11140359\unitlength}\raggedright \scriptsize{z}\end{minipage}}}%
    \put(0.35377013,0.88564859){\color[rgb]{0,0,0}\makebox(0,0)[lt]{\begin{minipage}{0.11140359\unitlength}\raggedright \scriptsize{y}\end{minipage}}}%
    \put(0.40893598,0.69485364){\color[rgb]{0,0,0}\makebox(0,0)[lt]{\begin{minipage}{0.11140359\unitlength}\raggedright \scriptsize{x}\end{minipage}}}%
    \put(0.19007441,0.98034389){\color[rgb]{0,0,0}\makebox(0,0)[lt]{\begin{minipage}{0.11140359\unitlength}\raggedright \scriptsize{z}\end{minipage}}}%
  \end{picture}%
\endgroup%

%% file: images/pdf/steady-state-entanglement.pdf_tex
\begingroup%
  \makeatletter%
  \providecommand\color[2][]{%
    \errmessage{(Inkscape) Color is used for the text in Inkscape, but the package 'color.sty' is not loaded}%
    \renewcommand\color[2][]{}%
  }%
  \providecommand\transparent[1]{%
    \errmessage{(Inkscape) Transparency is used (non-zero) for the text in Inkscape, but the package 'transparent.sty' is not loaded}%
    \renewcommand\transparent[1]{}%
  }%
  \providecommand\rotatebox[2]{#2}%
  \ifx\svgwidth\undefined%
    \setlength{\unitlength}{949.5369873bp}%
    \ifx\svgscale\undefined%
      \relax%
    \else%
      \setlength{\unitlength}{\unitlength * \real{\svgscale}}%
    \fi%
  \else%
    \setlength{\unitlength}{\svgwidth}%
  \fi%
  \global\let\svgwidth\undefined%
  \global\let\svgscale\undefined%
  \makeatother%
  \begin{picture}(1,0.87333934)%
    \put(0,0){\includegraphics[width=\unitlength,page=1]{steady-state-entanglement.pdf}}%
    \put(0.0739752,0.81133215){\color[rgb]{0,0,0}\makebox(0,0)[rb]{\smash{$-0.65$}}}%
    \put(0.0739752,0.70596167){\color[rgb]{0,0,0}\makebox(0,0)[rb]{\smash{$-0.70$}}}%
    \put(0.0739752,0.60059113){\color[rgb]{0,0,0}\makebox(0,0)[rb]{\smash{$-0.75$}}}%
    \put(0.0739752,0.49522064){\color[rgb]{0,0,0}\makebox(0,0)[rb]{\smash{$-0.80$}}}%
    \put(0.0739752,0.3898501){\color[rgb]{0,0,0}\makebox(0,0)[rb]{\smash{$-0.85$}}}%
    \put(0.0739752,0.28447957){\color[rgb]{0,0,0}\makebox(0,0)[rb]{\smash{$-0.90$}}}%
    \put(0.0739752,0.17910903){\color[rgb]{0,0,0}\makebox(0,0)[rb]{\smash{$-0.95$}}}%
    \put(0.0739752,0.07373849){\color[rgb]{0,0,0}\makebox(0,0)[rb]{\smash{$-1.00$}}}%
    \put(0.10504565,0.01808865){\color[rgb]{0,0,0}\makebox(0,0)[b]{\smash{$0.00$}}}%
    \put(0.25257019,0.01808865){\color[rgb]{0,0,0}\makebox(0,0)[b]{\smash{$0.02$}}}%
    \put(0.40009472,0.01808865){\color[rgb]{0,0,0}\makebox(0,0)[b]{\smash{$0.04$}}}%
    \put(0.54761931,0.01808865){\color[rgb]{0,0,0}\makebox(0,0)[b]{\smash{$0.06$}}}%
    \put(0.69514378,0.01808865){\color[rgb]{0,0,0}\makebox(0,0)[b]{\smash{$0.08$}}}%
    \put(0.8426683,0.01808865){\color[rgb]{0,0,0}\makebox(0,0)[b]{\smash{$0.10$}}}%
    \put(0,0){\includegraphics[width=\unitlength,page=2]{steady-state-entanglement.pdf}}%
    \put(0.87579839,0.05515936){\color[rgb]{0,0,0}\makebox(0,0)[lb]{\smash{$\delta t$}}}%
    \put(0.08922634,0.85008788){\color[rgb]{0,0,0}\makebox(0,0)[b]{\smash{$\eta$}}}%
    \put(0.91380079,0.69977012){\color[rgb]{0,0,0}\rotatebox{90}{\makebox(0,0)[b]{\smash{Concurrence}}}}%
    \put(0.96813829,0.80119647){\color[rgb]{0,0,0}\makebox(0,0)[lb]{\smash{$0$}}}%
    \put(0.96813829,0.73042513){\color[rgb]{0,0,0}\makebox(0,0)[lb]{\smash{$0.1$}}}%
    \put(0.96813829,0.65965377){\color[rgb]{0,0,0}\makebox(0,0)[lb]{\smash{$0.2$}}}%
    \put(0.96813829,0.58888246){\color[rgb]{0,0,0}\makebox(0,0)[lb]{\smash{$0.3$}}}%
    \put(0,0){\includegraphics[width=\unitlength,page=3]{steady-state-entanglement.pdf}}%
  \end{picture}%
\endgroup%

%% file: ms.bbl
\begin{thebibliography}{49}%
\makeatletter
\providecommand \@ifxundefined [1]{%
 \@ifx{#1\undefined}
}%
\providecommand \@ifnum [1]{%
 \ifnum #1\expandafter \@firstoftwo
 \else \expandafter \@secondoftwo
 \fi
}%
\providecommand \@ifx [1]{%
 \ifx #1\expandafter \@firstoftwo
 \else \expandafter \@secondoftwo
 \fi
}%
\providecommand \natexlab [1]{#1}%
\providecommand \enquote  [1]{``#1''}%
\providecommand \bibnamefont  [1]{#1}%
\providecommand \bibfnamefont [1]{#1}%
\providecommand \citenamefont [1]{#1}%
\providecommand \href@noop [0]{\@secondoftwo}%
\providecommand \href [0]{\begingroup \@sanitize@url \@href}%
\providecommand \@href[1]{\@@startlink{#1}\@@href}%
\providecommand \@@href[1]{\endgroup#1\@@endlink}%
\providecommand \@sanitize@url [0]{\catcode `\\12\catcode `\$12\catcode
  `\&12\catcode `\#12\catcode `\^12\catcode `\_12\catcode `\%12\relax}%
\providecommand \@@startlink[1]{}%
\providecommand \@@endlink[0]{}%
\providecommand \url  [0]{\begingroup\@sanitize@url \@url }%
\providecommand \@url [1]{\endgroup\@href {#1}{\urlprefix }}%
\providecommand \urlprefix  [0]{URL }%
\providecommand \Eprint [0]{\href }%
\providecommand \doibase [0]{http://dx.doi.org/}%
\providecommand \selectlanguage [0]{\@gobble}%
\providecommand \bibinfo  [0]{\@secondoftwo}%
\providecommand \bibfield  [0]{\@secondoftwo}%
\providecommand \translation [1]{[#1]}%
\providecommand \BibitemOpen [0]{}%
\providecommand \bibitemStop [0]{}%
\providecommand \bibitemNoStop [0]{.\EOS\space}%
\providecommand \EOS [0]{\spacefactor3000\relax}%
\providecommand \BibitemShut  [1]{\csname bibitem#1\endcsname}%
\let\auto@bib@innerbib\@empty
\bibitem [{\citenamefont {{Schr{\"o}dinger}}(1935)}]{schroedinger}%
  \BibitemOpen
  \bibfield  {author} {\bibinfo {author} {\bibfnamefont {E.}~\bibnamefont
  {{Schr{\"o}dinger}}},\ }\href {\doibase 10.1017/S0305004100013554} {\bibfield
   {journal} {\bibinfo  {journal} {Proceedings of the Cambridge Philosophical
  Society}\ }\textbf {\bibinfo {volume} {31}},\ \bibinfo {pages} {555}
  (\bibinfo {year} {1935})}\BibitemShut {NoStop}%
\bibitem [{\citenamefont {Einstein}\ \emph {et~al.}(1935)\citenamefont
  {Einstein}, \citenamefont {Podolsky},\ and\ \citenamefont {Rosen}}]{epr}%
  \BibitemOpen
  \bibfield  {author} {\bibinfo {author} {\bibfnamefont {A.}~\bibnamefont
  {Einstein}}, \bibinfo {author} {\bibfnamefont {B.}~\bibnamefont {Podolsky}},
  \ and\ \bibinfo {author} {\bibfnamefont {N.}~\bibnamefont {Rosen}},\ }\href
  {\doibase 10.1103/PhysRev.47.777} {\bibfield  {journal} {\bibinfo  {journal}
  {Phys. Rev.}\ }\textbf {\bibinfo {volume} {47}},\ \bibinfo {pages} {777}
  (\bibinfo {year} {1935})}\BibitemShut {NoStop}%
\bibitem [{\citenamefont {Bell}(1964)}]{Bell:1964kc}%
  \BibitemOpen
  \bibfield  {author} {\bibinfo {author} {\bibfnamefont {J.~S.}\ \bibnamefont
  {Bell}},\ }\href@noop {} {\bibfield  {journal} {\bibinfo  {journal}
  {Physics}\ }\textbf {\bibinfo {volume} {1}},\ \bibinfo {pages} {195}
  (\bibinfo {year} {1964})}\BibitemShut {NoStop}%
\bibitem [{\citenamefont {Wiseman}\ \emph {et~al.}(2007)\citenamefont
  {Wiseman}, \citenamefont {Jones},\ and\ \citenamefont
  {Doherty}}]{wiseman-steering}%
  \BibitemOpen
  \bibfield  {author} {\bibinfo {author} {\bibfnamefont {H.~M.}\ \bibnamefont
  {Wiseman}}, \bibinfo {author} {\bibfnamefont {S.~J.}\ \bibnamefont {Jones}},
  \ and\ \bibinfo {author} {\bibfnamefont {A.~C.}\ \bibnamefont {Doherty}},\
  }\href {\doibase 10.1103/PhysRevLett.98.140402} {\bibfield  {journal}
  {\bibinfo  {journal} {Phys. Rev. Lett.}\ }\textbf {\bibinfo {volume} {98}},\
  \bibinfo {pages} {140402} (\bibinfo {year} {2007})}\BibitemShut {NoStop}%
\bibitem [{\citenamefont {Jones}\ \emph {et~al.}(2007)\citenamefont {Jones},
  \citenamefont {Wiseman},\ and\ \citenamefont {Doherty}}]{jones}%
  \BibitemOpen
  \bibfield  {author} {\bibinfo {author} {\bibfnamefont {S.~J.}\ \bibnamefont
  {Jones}}, \bibinfo {author} {\bibfnamefont {H.~M.}\ \bibnamefont {Wiseman}},
  \ and\ \bibinfo {author} {\bibfnamefont {A.~C.}\ \bibnamefont {Doherty}},\
  }\href {\doibase 10.1103/PhysRevA.76.052116} {\bibfield  {journal} {\bibinfo
  {journal} {Phys. Rev. A}\ }\textbf {\bibinfo {volume} {76}},\ \bibinfo
  {pages} {052116} (\bibinfo {year} {2007})}\BibitemShut {NoStop}%
\bibitem [{\citenamefont {Cavalcanti}\ \emph {et~al.}(2009)\citenamefont
  {Cavalcanti}, \citenamefont {Jones}, \citenamefont {Wiseman},\ and\
  \citenamefont {Reid}}]{steering-criteria}%
  \BibitemOpen
  \bibfield  {author} {\bibinfo {author} {\bibfnamefont {E.~G.}\ \bibnamefont
  {Cavalcanti}}, \bibinfo {author} {\bibfnamefont {S.~J.}\ \bibnamefont
  {Jones}}, \bibinfo {author} {\bibfnamefont {H.~M.}\ \bibnamefont {Wiseman}},
  \ and\ \bibinfo {author} {\bibfnamefont {M.~D.}\ \bibnamefont {Reid}},\
  }\href {\doibase 10.1103/PhysRevA.80.032112} {\bibfield  {journal} {\bibinfo
  {journal} {Phys. Rev. A}\ }\textbf {\bibinfo {volume} {80}},\ \bibinfo
  {pages} {032112} (\bibinfo {year} {2009})}\BibitemShut {NoStop}%
\bibitem [{\citenamefont {Costa}\ and\ \citenamefont
  {Angelo}(2016)}]{costa-steering}%
  \BibitemOpen
  \bibfield  {author} {\bibinfo {author} {\bibfnamefont {A.~C.~S.}\
  \bibnamefont {Costa}}\ and\ \bibinfo {author} {\bibfnamefont {R.~M.}\
  \bibnamefont {Angelo}},\ }\href {\doibase 10.1103/PhysRevA.93.020103}
  {\bibfield  {journal} {\bibinfo  {journal} {Phys. Rev. A}\ }\textbf {\bibinfo
  {volume} {93}},\ \bibinfo {pages} {020103} (\bibinfo {year}
  {2016})}\BibitemShut {NoStop}%
\bibitem [{\citenamefont {Costa}\ \emph {et~al.}(2016)\citenamefont {Costa},
  \citenamefont {Beims},\ and\ \citenamefont {Angelo}}]{costa-channels}%
  \BibitemOpen
  \bibfield  {author} {\bibinfo {author} {\bibfnamefont {A.}~\bibnamefont
  {Costa}}, \bibinfo {author} {\bibfnamefont {M.}~\bibnamefont {Beims}}, \ and\
  \bibinfo {author} {\bibfnamefont {R.}~\bibnamefont {Angelo}},\ }\href
  {\doibase https://doi.org/10.1016/j.physa.2016.05.068} {\bibfield  {journal}
  {\bibinfo  {journal} {Physica A: Statistical Mechanics and its Applications}\
  }\textbf {\bibinfo {volume} {461}},\ \bibinfo {pages} {469 } (\bibinfo {year}
  {2016})}\BibitemShut {NoStop}%
\bibitem [{\citenamefont {Nguyen}\ and\ \citenamefont
  {Vu}(2016)}]{nguyen-geometry}%
  \BibitemOpen
  \bibfield  {author} {\bibinfo {author} {\bibfnamefont {H.~C.}\ \bibnamefont
  {Nguyen}}\ and\ \bibinfo {author} {\bibfnamefont {T.}~\bibnamefont {Vu}},\
  }\href {\doibase 10.1103/PhysRevA.94.012114} {\bibfield  {journal} {\bibinfo
  {journal} {Phys. Rev. A}\ }\textbf {\bibinfo {volume} {94}},\ \bibinfo
  {pages} {012114} (\bibinfo {year} {2016})}\BibitemShut {NoStop}%
\bibitem [{\citenamefont {Uola}\ \emph {et~al.}(2015)\citenamefont {Uola},
  \citenamefont {Budroni}, \citenamefont {G\"uhne},\ and\ \citenamefont
  {Pellonp\"a\"a}}]{PhysRevLett.115.230402}%
  \BibitemOpen
  \bibfield  {author} {\bibinfo {author} {\bibfnamefont {R.}~\bibnamefont
  {Uola}}, \bibinfo {author} {\bibfnamefont {C.}~\bibnamefont {Budroni}},
  \bibinfo {author} {\bibfnamefont {O.}~\bibnamefont {G\"uhne}}, \ and\
  \bibinfo {author} {\bibfnamefont {J.-P.}\ \bibnamefont {Pellonp\"a\"a}},\
  }\href {\doibase 10.1103/PhysRevLett.115.230402} {\bibfield  {journal}
  {\bibinfo  {journal} {Phys. Rev. Lett.}\ }\textbf {\bibinfo {volume} {115}},\
  \bibinfo {pages} {230402} (\bibinfo {year} {2015})}\BibitemShut {NoStop}%
\bibitem [{\citenamefont {Uola}\ \emph {et~al.}(2014)\citenamefont {Uola},
  \citenamefont {Moroder},\ and\ \citenamefont
  {G\"uhne}}]{PhysRevLett.113.160403}%
  \BibitemOpen
  \bibfield  {author} {\bibinfo {author} {\bibfnamefont {R.}~\bibnamefont
  {Uola}}, \bibinfo {author} {\bibfnamefont {T.}~\bibnamefont {Moroder}}, \
  and\ \bibinfo {author} {\bibfnamefont {O.}~\bibnamefont {G\"uhne}},\ }\href
  {\doibase 10.1103/PhysRevLett.113.160403} {\bibfield  {journal} {\bibinfo
  {journal} {Phys. Rev. Lett.}\ }\textbf {\bibinfo {volume} {113}},\ \bibinfo
  {pages} {160403} (\bibinfo {year} {2014})}\BibitemShut {NoStop}%
\bibitem [{\citenamefont {Skrzypczyk}\ \emph {et~al.}(2014)\citenamefont
  {Skrzypczyk}, \citenamefont {Navascu\'es},\ and\ \citenamefont
  {Cavalcanti}}]{PhysRevLett.112.180404}%
  \BibitemOpen
  \bibfield  {author} {\bibinfo {author} {\bibfnamefont {P.}~\bibnamefont
  {Skrzypczyk}}, \bibinfo {author} {\bibfnamefont {M.}~\bibnamefont
  {Navascu\'es}}, \ and\ \bibinfo {author} {\bibfnamefont {D.}~\bibnamefont
  {Cavalcanti}},\ }\href {\doibase 10.1103/PhysRevLett.112.180404} {\bibfield
  {journal} {\bibinfo  {journal} {Phys. Rev. Lett.}\ }\textbf {\bibinfo
  {volume} {112}},\ \bibinfo {pages} {180404} (\bibinfo {year}
  {2014})}\BibitemShut {NoStop}%
\bibitem [{\citenamefont {Quintino}\ \emph {et~al.}(2014)\citenamefont
  {Quintino}, \citenamefont {V\'ertesi},\ and\ \citenamefont
  {Brunner}}]{PhysRevLett.113.160402}%
  \BibitemOpen
  \bibfield  {author} {\bibinfo {author} {\bibfnamefont {M.~T.}\ \bibnamefont
  {Quintino}}, \bibinfo {author} {\bibfnamefont {T.}~\bibnamefont {V\'ertesi}},
  \ and\ \bibinfo {author} {\bibfnamefont {N.}~\bibnamefont {Brunner}},\ }\href
  {\doibase 10.1103/PhysRevLett.113.160402} {\bibfield  {journal} {\bibinfo
  {journal} {Phys. Rev. Lett.}\ }\textbf {\bibinfo {volume} {113}},\ \bibinfo
  {pages} {160402} (\bibinfo {year} {2014})}\BibitemShut {NoStop}%
\bibitem [{\citenamefont {Saunders}\ \emph {et~al.}(2010)\citenamefont
  {Saunders}, \citenamefont {Pryde}, \citenamefont {Wiseman},\ and\
  \citenamefont {Jones}}]{saunders2010}%
  \BibitemOpen
  \bibfield  {author} {\bibinfo {author} {\bibfnamefont {D.~J.}\ \bibnamefont
  {Saunders}}, \bibinfo {author} {\bibfnamefont {G.~J.}\ \bibnamefont {Pryde}},
  \bibinfo {author} {\bibfnamefont {H.~M.}\ \bibnamefont {Wiseman}}, \ and\
  \bibinfo {author} {\bibfnamefont {S.~J.}\ \bibnamefont {Jones}},\ }\href
  {\doibase 10.1038/nphys1766} {\bibfield  {journal} {\bibinfo  {journal}
  {Nature Physics}\ }\textbf {\bibinfo {volume} {6}},\ \bibinfo {pages} {845}
  (\bibinfo {year} {2010})}\BibitemShut {NoStop}%
\bibitem [{\citenamefont {Wittmann}\ \emph {et~al.}(2012)\citenamefont
  {Wittmann}, \citenamefont {Ramelow}, \citenamefont {Steinlechner},
  \citenamefont {Langford}, \citenamefont {Brunner}, \citenamefont {Wiseman},
  \citenamefont {Ursin},\ and\ \citenamefont {Zeilinger}}]{wittman2012}%
  \BibitemOpen
  \bibfield  {author} {\bibinfo {author} {\bibfnamefont {B.}~\bibnamefont
  {Wittmann}}, \bibinfo {author} {\bibfnamefont {S.}~\bibnamefont {Ramelow}},
  \bibinfo {author} {\bibfnamefont {F.}~\bibnamefont {Steinlechner}}, \bibinfo
  {author} {\bibfnamefont {N.~K.}\ \bibnamefont {Langford}}, \bibinfo {author}
  {\bibfnamefont {N.}~\bibnamefont {Brunner}}, \bibinfo {author} {\bibfnamefont
  {H.~M.}\ \bibnamefont {Wiseman}}, \bibinfo {author} {\bibfnamefont
  {R.}~\bibnamefont {Ursin}}, \ and\ \bibinfo {author} {\bibfnamefont
  {A.}~\bibnamefont {Zeilinger}},\ }\href
  {http://stacks.iop.org/1367-2630/14/i=5/a=053030} {\bibfield  {journal}
  {\bibinfo  {journal} {New Journal of Physics}\ }\textbf {\bibinfo {volume}
  {14}},\ \bibinfo {pages} {053030} (\bibinfo {year} {2012})}\BibitemShut
  {NoStop}%
\bibitem [{\citenamefont {Wollmann}\ \emph {et~al.}(2016)\citenamefont
  {Wollmann}, \citenamefont {Walk}, \citenamefont {Bennet}, \citenamefont
  {Wiseman},\ and\ \citenamefont {Pryde}}]{PhysRevLett.116.160403}%
  \BibitemOpen
  \bibfield  {author} {\bibinfo {author} {\bibfnamefont {S.}~\bibnamefont
  {Wollmann}}, \bibinfo {author} {\bibfnamefont {N.}~\bibnamefont {Walk}},
  \bibinfo {author} {\bibfnamefont {A.~J.}\ \bibnamefont {Bennet}}, \bibinfo
  {author} {\bibfnamefont {H.~M.}\ \bibnamefont {Wiseman}}, \ and\ \bibinfo
  {author} {\bibfnamefont {G.~J.}\ \bibnamefont {Pryde}},\ }\href {\doibase
  10.1103/PhysRevLett.116.160403} {\bibfield  {journal} {\bibinfo  {journal}
  {Phys. Rev. Lett.}\ }\textbf {\bibinfo {volume} {116}},\ \bibinfo {pages}
  {160403} (\bibinfo {year} {2016})}\BibitemShut {NoStop}%
\bibitem [{\citenamefont {Xiao}\ \emph {et~al.}(2017)\citenamefont {Xiao},
  \citenamefont {Ye}, \citenamefont {Sun}, \citenamefont {Xu}, \citenamefont
  {Li},\ and\ \citenamefont {Guo}}]{PhysRevLett.118.140404}%
  \BibitemOpen
  \bibfield  {author} {\bibinfo {author} {\bibfnamefont {Y.}~\bibnamefont
  {Xiao}}, \bibinfo {author} {\bibfnamefont {X.-J.}\ \bibnamefont {Ye}},
  \bibinfo {author} {\bibfnamefont {K.}~\bibnamefont {Sun}}, \bibinfo {author}
  {\bibfnamefont {J.-S.}\ \bibnamefont {Xu}}, \bibinfo {author} {\bibfnamefont
  {C.-F.}\ \bibnamefont {Li}}, \ and\ \bibinfo {author} {\bibfnamefont {G.-C.}\
  \bibnamefont {Guo}},\ }\href {\doibase 10.1103/PhysRevLett.118.140404}
  {\bibfield  {journal} {\bibinfo  {journal} {Phys. Rev. Lett.}\ }\textbf
  {\bibinfo {volume} {118}},\ \bibinfo {pages} {140404} (\bibinfo {year}
  {2017})}\BibitemShut {NoStop}%
\bibitem [{\citenamefont {Gisin}(1991)}]{gisin}%
  \BibitemOpen
  \bibfield  {author} {\bibinfo {author} {\bibfnamefont {N.}~\bibnamefont
  {Gisin}},\ }\href {\doibase http://dx.doi.org/10.1016/0375-9601(91)90805-I}
  {\bibfield  {journal} {\bibinfo  {journal} {Physics Letters A}\ }\textbf
  {\bibinfo {volume} {154}},\ \bibinfo {pages} {201 } (\bibinfo {year}
  {1991})}\BibitemShut {NoStop}%
\bibitem [{\citenamefont {Wiseman}\ and\ \citenamefont
  {Gambetta}(2012)}]{wiseman-gambetta}%
  \BibitemOpen
  \bibfield  {author} {\bibinfo {author} {\bibfnamefont {H.~M.}\ \bibnamefont
  {Wiseman}}\ and\ \bibinfo {author} {\bibfnamefont {J.~M.}\ \bibnamefont
  {Gambetta}},\ }\href {\doibase 10.1103/PhysRevLett.108.220402} {\bibfield
  {journal} {\bibinfo  {journal} {Phys. Rev. Lett.}\ }\textbf {\bibinfo
  {volume} {108}},\ \bibinfo {pages} {220402} (\bibinfo {year}
  {2012})}\BibitemShut {NoStop}%
\bibitem [{\citenamefont {Daryanoosh}\ and\ \citenamefont
  {Wiseman}(2014)}]{daryanoosh_quantum_2014}%
  \BibitemOpen
  \bibfield  {author} {\bibinfo {author} {\bibfnamefont {S.}~\bibnamefont
  {Daryanoosh}}\ and\ \bibinfo {author} {\bibfnamefont {H.~M.}\ \bibnamefont
  {Wiseman}},\ }\href {\doibase 10.1088/1367-2630/16/6/063028} {\bibfield
  {journal} {\bibinfo  {journal} {New Journal of Physics}\ }\textbf {\bibinfo
  {volume} {16}},\ \bibinfo {pages} {063028} (\bibinfo {year}
  {2014})}\BibitemShut {NoStop}%
\bibitem [{\citenamefont {Daryanoosh}\ \emph {et~al.}(2015)\citenamefont
  {Daryanoosh}, \citenamefont {Wiseman},\ and\ \citenamefont
  {Gambetta}}]{daryanoosh_detector_2015}%
  \BibitemOpen
  \bibfield  {author} {\bibinfo {author} {\bibfnamefont {S.}~\bibnamefont
  {Daryanoosh}}, \bibinfo {author} {\bibfnamefont {H.~M.}\ \bibnamefont
  {Wiseman}}, \ and\ \bibinfo {author} {\bibfnamefont {J.~M.}\ \bibnamefont
  {Gambetta}},\ }\href {\doibase 10.1103/PhysRevA.92.042114} {\bibfield
  {journal} {\bibinfo  {journal} {Physical Review A}\ }\textbf {\bibinfo
  {volume} {92}},\ \bibinfo {pages} {042114} (\bibinfo {year}
  {2015})}\BibitemShut {NoStop}%
\bibitem [{\citenamefont {Gorini}\ \emph {et~al.}(1976)\citenamefont {Gorini},
  \citenamefont {Kossakowski},\ and\ \citenamefont
  {Sudarshan}}]{GoriniKossakowskiSudarshan}%
  \BibitemOpen
  \bibfield  {author} {\bibinfo {author} {\bibfnamefont {V.}~\bibnamefont
  {Gorini}}, \bibinfo {author} {\bibfnamefont {A.}~\bibnamefont {Kossakowski}},
  \ and\ \bibinfo {author} {\bibfnamefont {E.~C.~G.}\ \bibnamefont
  {Sudarshan}},\ }\href {\doibase 10.1063/1.522979} {\bibfield  {journal}
  {\bibinfo  {journal} {Journal of Mathematical Physics}\ }\textbf {\bibinfo
  {volume} {17}},\ \bibinfo {pages} {821} (\bibinfo {year} {1976})}\BibitemShut
  {NoStop}%
\bibitem [{\citenamefont {Lindblad}(1976)}]{lindblad}%
  \BibitemOpen
  \bibfield  {author} {\bibinfo {author} {\bibfnamefont {G.}~\bibnamefont
  {Lindblad}},\ }\href {\doibase 10.1007/BF01608499} {\bibfield  {journal}
  {\bibinfo  {journal} {Communications in Mathematical Physics}\ }\textbf
  {\bibinfo {volume} {48}},\ \bibinfo {pages} {119} (\bibinfo {year}
  {1976})}\BibitemShut {NoStop}%
\bibitem [{\citenamefont {Carmichael}(1993)}]{carmichael}%
  \BibitemOpen
  \bibfield  {author} {\bibinfo {author} {\bibfnamefont {H.}~\bibnamefont
  {Carmichael}},\ }\href {https://cds.cern.ch/record/1631392} {\emph {\bibinfo
  {title} {{An open systems approach to quantum optics}}}},\ Lecture Notes in
  Physics Monographs\ (\bibinfo  {publisher} {Springer},\ \bibinfo {address}
  {Berlin},\ \bibinfo {year} {1993})\BibitemShut {NoStop}%
\bibitem [{\citenamefont {Wiseman}\ and\ \citenamefont
  {Milburn}(2009)}]{wiseman-milburn}%
  \BibitemOpen
  \bibfield  {author} {\bibinfo {author} {\bibfnamefont {H.~M.}\ \bibnamefont
  {Wiseman}}\ and\ \bibinfo {author} {\bibfnamefont {G.~J.}\ \bibnamefont
  {Milburn}},\ }\href {https://cds.cern.ch/record/1320323} {\emph {\bibinfo
  {title} {{Quantum Measurement and Control}}}}\ (\bibinfo  {publisher}
  {Cambridge University Press},\ \bibinfo {address} {Leiden},\ \bibinfo {year}
  {2009})\BibitemShut {NoStop}%
\bibitem [{\citenamefont {Wiseman}\ and\ \citenamefont {Brady}(2000)}]{robust}%
  \BibitemOpen
  \bibfield  {author} {\bibinfo {author} {\bibfnamefont {H.~M.}\ \bibnamefont
  {Wiseman}}\ and\ \bibinfo {author} {\bibfnamefont {Z.}~\bibnamefont
  {Brady}},\ }\href {\doibase 10.1103/PhysRevA.62.023805} {\bibfield  {journal}
  {\bibinfo  {journal} {Phys. Rev. A}\ }\textbf {\bibinfo {volume} {62}},\
  \bibinfo {pages} {023805} (\bibinfo {year} {2000})}\BibitemShut {NoStop}%
\bibitem [{\citenamefont {Ziman}\ \emph {et~al.}(2005)\citenamefont {Ziman},
  \citenamefont {\v{S}telmachovi\v{c}},\ and\ \citenamefont
  {Bu\v{z}ek}}]{ziman-collision}%
  \BibitemOpen
  \bibfield  {author} {\bibinfo {author} {\bibfnamefont {M.}~\bibnamefont
  {Ziman}}, \bibinfo {author} {\bibfnamefont {P.}~\bibnamefont
  {\v{S}telmachovi\v{c}}}, \ and\ \bibinfo {author} {\bibfnamefont
  {V.}~\bibnamefont {Bu\v{z}ek}},\ }\href {\doibase 10.1007/s11080-005-0488-0}
  {\bibfield  {journal} {\bibinfo  {journal} {Open Systems \& Information
  Dynamics}\ }\textbf {\bibinfo {volume} {12}},\ \bibinfo {pages} {81}
  (\bibinfo {year} {2005})}\BibitemShut {NoStop}%
\bibitem [{\citenamefont {Ziman}\ and\ \citenamefont {Bu\ifmmode~\check{z}\else
  \v{z}\fi{}ek}(2005)}]{ziman-decoherences}%
  \BibitemOpen
  \bibfield  {author} {\bibinfo {author} {\bibfnamefont {M.}~\bibnamefont
  {Ziman}}\ and\ \bibinfo {author} {\bibfnamefont {V.}~\bibnamefont
  {Bu\ifmmode~\check{z}\else \v{z}\fi{}ek}},\ }\href {\doibase
  10.1103/PhysRevA.72.022110} {\bibfield  {journal} {\bibinfo  {journal} {Phys.
  Rev. A}\ }\textbf {\bibinfo {volume} {72}},\ \bibinfo {pages} {022110}
  (\bibinfo {year} {2005})}\BibitemShut {NoStop}%
\bibitem [{\citenamefont {Ziman}\ \emph {et~al.}(2002)\citenamefont {Ziman},
  \citenamefont {\ifmmode \check{S}\else
  \v{S}\fi{}telmachovi\ifmmode~\check{c}\else \v{c}\fi{}}, \citenamefont
  {Bu\ifmmode~\check{z}\else \v{z}\fi{}ek}, \citenamefont {Hillery},
  \citenamefont {Scarani},\ and\ \citenamefont {Gisin}}]{ziman-diluting}%
  \BibitemOpen
  \bibfield  {author} {\bibinfo {author} {\bibfnamefont {M.}~\bibnamefont
  {Ziman}}, \bibinfo {author} {\bibfnamefont {P.}~\bibnamefont {\ifmmode
  \check{S}\else \v{S}\fi{}telmachovi\ifmmode~\check{c}\else \v{c}\fi{}}},
  \bibinfo {author} {\bibfnamefont {V.}~\bibnamefont {Bu\ifmmode~\check{z}\else
  \v{z}\fi{}ek}}, \bibinfo {author} {\bibfnamefont {M.}~\bibnamefont
  {Hillery}}, \bibinfo {author} {\bibfnamefont {V.}~\bibnamefont {Scarani}}, \
  and\ \bibinfo {author} {\bibfnamefont {N.}~\bibnamefont {Gisin}},\ }\href
  {\doibase 10.1103/PhysRevA.65.042105} {\bibfield  {journal} {\bibinfo
  {journal} {Phys. Rev. A}\ }\textbf {\bibinfo {volume} {65}},\ \bibinfo
  {pages} {042105} (\bibinfo {year} {2002})}\BibitemShut {NoStop}%
\bibitem [{\citenamefont {Scarani}\ \emph {et~al.}(2002)\citenamefont
  {Scarani}, \citenamefont {Ziman}, \citenamefont {\ifmmode \check{S}\else
  \v{S}\fi{}telmachovi\ifmmode~\check{c}\else \v{c}\fi{}}, \citenamefont
  {Gisin},\ and\ \citenamefont {Bu\ifmmode~\check{z}\else
  \v{z}\fi{}ek}}]{scarani-thermalizing}%
  \BibitemOpen
  \bibfield  {author} {\bibinfo {author} {\bibfnamefont {V.}~\bibnamefont
  {Scarani}}, \bibinfo {author} {\bibfnamefont {M.}~\bibnamefont {Ziman}},
  \bibinfo {author} {\bibfnamefont {P.}~\bibnamefont {\ifmmode \check{S}\else
  \v{S}\fi{}telmachovi\ifmmode~\check{c}\else \v{c}\fi{}}}, \bibinfo {author}
  {\bibfnamefont {N.}~\bibnamefont {Gisin}}, \ and\ \bibinfo {author}
  {\bibfnamefont {V.}~\bibnamefont {Bu\ifmmode~\check{z}\else \v{z}\fi{}ek}},\
  }\href {\doibase 10.1103/PhysRevLett.88.097905} {\bibfield  {journal}
  {\bibinfo  {journal} {Phys. Rev. Lett.}\ }\textbf {\bibinfo {volume} {88}},\
  \bibinfo {pages} {097905} (\bibinfo {year} {2002})}\BibitemShut {NoStop}%
\bibitem [{\citenamefont {Giovannetti}\ and\ \citenamefont
  {Palma}(2012{\natexlab{a}})}]{giovannetti-collision}%
  \BibitemOpen
  \bibfield  {author} {\bibinfo {author} {\bibfnamefont {V.}~\bibnamefont
  {Giovannetti}}\ and\ \bibinfo {author} {\bibfnamefont {G.~M.}\ \bibnamefont
  {Palma}},\ }\href {http://stacks.iop.org/0953-4075/45/i=15/a=154003}
  {\bibfield  {journal} {\bibinfo  {journal} {Journal of Physics B: Atomic,
  Molecular and Optical Physics}\ }\textbf {\bibinfo {volume} {45}},\ \bibinfo
  {pages} {154003} (\bibinfo {year} {2012}{\natexlab{a}})}\BibitemShut
  {NoStop}%
\bibitem [{\citenamefont {Giovannetti}\ and\ \citenamefont
  {Palma}(2012{\natexlab{b}})}]{giovannetti-correlated}%
  \BibitemOpen
  \bibfield  {author} {\bibinfo {author} {\bibfnamefont {V.}~\bibnamefont
  {Giovannetti}}\ and\ \bibinfo {author} {\bibfnamefont {G.~M.}\ \bibnamefont
  {Palma}},\ }\href {\doibase 10.1103/PhysRevLett.108.040401} {\bibfield
  {journal} {\bibinfo  {journal} {Phys. Rev. Lett.}\ }\textbf {\bibinfo
  {volume} {108}},\ \bibinfo {pages} {040401} (\bibinfo {year}
  {2012}{\natexlab{b}})}\BibitemShut {NoStop}%
\bibitem [{\citenamefont {Filippov}\ \emph {et~al.}(2017)\citenamefont
  {Filippov}, \citenamefont {Piilo}, \citenamefont {Maniscalco},\ and\
  \citenamefont {Ziman}}]{filippov2017}%
  \BibitemOpen
  \bibfield  {author} {\bibinfo {author} {\bibfnamefont {S.~N.}\ \bibnamefont
  {Filippov}}, \bibinfo {author} {\bibfnamefont {J.}~\bibnamefont {Piilo}},
  \bibinfo {author} {\bibfnamefont {S.}~\bibnamefont {Maniscalco}}, \ and\
  \bibinfo {author} {\bibfnamefont {M.}~\bibnamefont {Ziman}},\ }\href
  {\doibase 10.1103/PhysRevA.96.032111} {\bibfield  {journal} {\bibinfo
  {journal} {Phys. Rev. A}\ }\textbf {\bibinfo {volume} {96}},\ \bibinfo
  {pages} {032111} (\bibinfo {year} {2017})}\BibitemShut {NoStop}%
\bibitem [{\citenamefont {Lorenzo}\ \emph {et~al.}(2017)\citenamefont
  {Lorenzo}, \citenamefont {Ciccarello},\ and\ \citenamefont
  {Palma}}]{composite-collision}%
  \BibitemOpen
  \bibfield  {author} {\bibinfo {author} {\bibfnamefont {S.}~\bibnamefont
  {Lorenzo}}, \bibinfo {author} {\bibfnamefont {F.}~\bibnamefont {Ciccarello}},
  \ and\ \bibinfo {author} {\bibfnamefont {G.~M.}\ \bibnamefont {Palma}},\
  }\href {\doibase 10.1103/PhysRevA.96.032107} {\bibfield  {journal} {\bibinfo
  {journal} {Phys. Rev. A}\ }\textbf {\bibinfo {volume} {96}},\ \bibinfo
  {pages} {032107} (\bibinfo {year} {2017})}\BibitemShut {NoStop}%
\bibitem [{\citenamefont {Lorenzo}\ \emph {et~al.}(2016)\citenamefont
  {Lorenzo}, \citenamefont {Ciccarello},\ and\ \citenamefont
  {Palma}}]{memory-kernel}%
  \BibitemOpen
  \bibfield  {author} {\bibinfo {author} {\bibfnamefont {S.}~\bibnamefont
  {Lorenzo}}, \bibinfo {author} {\bibfnamefont {F.}~\bibnamefont {Ciccarello}},
  \ and\ \bibinfo {author} {\bibfnamefont {G.~M.}\ \bibnamefont {Palma}},\
  }\href {\doibase 10.1103/PhysRevA.93.052111} {\bibfield  {journal} {\bibinfo
  {journal} {Phys. Rev. A}\ }\textbf {\bibinfo {volume} {93}},\ \bibinfo
  {pages} {052111} (\bibinfo {year} {2016})}\BibitemShut {NoStop}%
\bibitem [{\citenamefont {Kretschmer}\ \emph {et~al.}(2016)\citenamefont
  {Kretschmer}, \citenamefont {Luoma},\ and\ \citenamefont
  {Strunz}}]{kretschmer}%
  \BibitemOpen
  \bibfield  {author} {\bibinfo {author} {\bibfnamefont {S.}~\bibnamefont
  {Kretschmer}}, \bibinfo {author} {\bibfnamefont {K.}~\bibnamefont {Luoma}}, \
  and\ \bibinfo {author} {\bibfnamefont {W.~T.}\ \bibnamefont {Strunz}},\
  }\href {\doibase 10.1103/PhysRevA.94.012106} {\bibfield  {journal} {\bibinfo
  {journal} {Phys. Rev. A}\ }\textbf {\bibinfo {volume} {94}},\ \bibinfo
  {pages} {012106} (\bibinfo {year} {2016})}\BibitemShut {NoStop}%
\bibitem [{\citenamefont {Luoma}\ \emph {et~al.}(2014)\citenamefont {Luoma},
  \citenamefont {Haikka},\ and\ \citenamefont {Piilo}}]{luoma-nm}%
  \BibitemOpen
  \bibfield  {author} {\bibinfo {author} {\bibfnamefont {K.}~\bibnamefont
  {Luoma}}, \bibinfo {author} {\bibfnamefont {P.}~\bibnamefont {Haikka}}, \
  and\ \bibinfo {author} {\bibfnamefont {J.}~\bibnamefont {Piilo}},\ }\href
  {\doibase 10.1103/PhysRevA.90.054101} {\bibfield  {journal} {\bibinfo
  {journal} {Phys. Rev. A}\ }\textbf {\bibinfo {volume} {90}},\ \bibinfo
  {pages} {054101} (\bibinfo {year} {2014})}\BibitemShut {NoStop}%
\bibitem [{\citenamefont {Ryb\'{a}r}\ \emph {et~al.}(2012)\citenamefont
  {Ryb\'{a}r}, \citenamefont {Filippov}, \citenamefont {Ziman},\ and\
  \citenamefont {Bu\v{z}ek}}]{rybar}%
  \BibitemOpen
  \bibfield  {author} {\bibinfo {author} {\bibfnamefont {T.}~\bibnamefont
  {Ryb\'{a}r}}, \bibinfo {author} {\bibfnamefont {S.~N.}\ \bibnamefont
  {Filippov}}, \bibinfo {author} {\bibfnamefont {M.}~\bibnamefont {Ziman}}, \
  and\ \bibinfo {author} {\bibfnamefont {V.}~\bibnamefont {Bu\v{z}ek}},\ }\href
  {http://stacks.iop.org/0953-4075/45/i=15/a=154006} {\bibfield  {journal}
  {\bibinfo  {journal} {Journal of Physics B: Atomic, Molecular and Optical
  Physics}\ }\textbf {\bibinfo {volume} {45}},\ \bibinfo {pages} {154006}
  (\bibinfo {year} {2012})}\BibitemShut {NoStop}%
\bibitem [{\citenamefont {Ciccarello}\ \emph {et~al.}(2013)\citenamefont
  {Ciccarello}, \citenamefont {Palma},\ and\ \citenamefont
  {Giovannetti}}]{ciccarello-nm}%
  \BibitemOpen
  \bibfield  {author} {\bibinfo {author} {\bibfnamefont {F.}~\bibnamefont
  {Ciccarello}}, \bibinfo {author} {\bibfnamefont {G.~M.}\ \bibnamefont
  {Palma}}, \ and\ \bibinfo {author} {\bibfnamefont {V.}~\bibnamefont
  {Giovannetti}},\ }\href {\doibase 10.1103/PhysRevA.87.040103} {\bibfield
  {journal} {\bibinfo  {journal} {Phys. Rev. A}\ }\textbf {\bibinfo {volume}
  {87}},\ \bibinfo {pages} {040103} (\bibinfo {year} {2013})}\BibitemShut
  {NoStop}%
\bibitem [{\citenamefont {\ifmmode~\mbox{\c{C}}\else \c{C}\fi{}akmak}\ \emph
  {et~al.}(2017)\citenamefont {\ifmmode~\mbox{\c{C}}\else \c{C}\fi{}akmak},
  \citenamefont {Pezzutto}, \citenamefont {Paternostro},\ and\ \citenamefont
  {M\"ustecapl\ifmmode \imath \else \i \fi{}o\ifmmode~\breve{g}\else
  \u{g}\fi{}lu}}]{cakmak-collision}%
  \BibitemOpen
  \bibfield  {author} {\bibinfo {author} {\bibfnamefont {B.}~\bibnamefont
  {\ifmmode~\mbox{\c{C}}\else \c{C}\fi{}akmak}}, \bibinfo {author}
  {\bibfnamefont {M.}~\bibnamefont {Pezzutto}}, \bibinfo {author}
  {\bibfnamefont {M.}~\bibnamefont {Paternostro}}, \ and\ \bibinfo {author}
  {\bibfnamefont {O.~E.}\ \bibnamefont {M\"ustecapl\ifmmode \imath \else \i
  \fi{}o\ifmmode~\breve{g}\else \u{g}\fi{}lu}},\ }\href {\doibase
  10.1103/PhysRevA.96.022109} {\bibfield  {journal} {\bibinfo  {journal} {Phys.
  Rev. A}\ }\textbf {\bibinfo {volume} {96}},\ \bibinfo {pages} {022109}
  (\bibinfo {year} {2017})}\BibitemShut {NoStop}%
\bibitem [{\citenamefont {Brun}(2002)}]{brun}%
  \BibitemOpen
  \bibfield  {author} {\bibinfo {author} {\bibfnamefont {T.~A.}\ \bibnamefont
  {Brun}},\ }\href {\doibase 10.1119/1.1475328} {\bibfield  {journal} {\bibinfo
   {journal} {American Journal of Physics}\ }\textbf {\bibinfo {volume} {70}},\
  \bibinfo {pages} {719} (\bibinfo {year} {2002})}\BibitemShut {NoStop}%
\bibitem [{\citenamefont {Nielsen}\ and\ \citenamefont
  {Chuang}(2000)}]{nielsen}%
  \BibitemOpen
  \bibfield  {author} {\bibinfo {author} {\bibfnamefont {M.}~\bibnamefont
  {Nielsen}}\ and\ \bibinfo {author} {\bibfnamefont {I.}~\bibnamefont
  {Chuang}},\ }\href@noop {} {\emph {\bibinfo {title} {Quantum Computation and
  Quantum Information}}},\ Cambridge Series on Information and the Natural
  Sciences\ (\bibinfo  {publisher} {Cambridge University Press},\ \bibinfo
  {year} {2000})\BibitemShut {NoStop}%
\bibitem [{\citenamefont {Wiseman}\ and\ \citenamefont
  {Toombes}(1999)}]{wiseman_quantum_1999}%
  \BibitemOpen
  \bibfield  {author} {\bibinfo {author} {\bibfnamefont {H.~M.}\ \bibnamefont
  {Wiseman}}\ and\ \bibinfo {author} {\bibfnamefont {G.~E.}\ \bibnamefont
  {Toombes}},\ }\href {\doibase 10.1103/PhysRevA.60.2474} {\bibfield  {journal}
  {\bibinfo  {journal} {Physical Review A}\ }\textbf {\bibinfo {volume} {60}},\
  \bibinfo {pages} {2474} (\bibinfo {year} {1999})}\BibitemShut {NoStop}%
\bibitem [{\citenamefont {Karasik}\ and\ \citenamefont
  {Wiseman}(2011)}]{karasik_how_2011}%
  \BibitemOpen
  \bibfield  {author} {\bibinfo {author} {\bibfnamefont {R.~I.}\ \bibnamefont
  {Karasik}}\ and\ \bibinfo {author} {\bibfnamefont {H.~M.}\ \bibnamefont
  {Wiseman}},\ }\href {\doibase 10.1103/PhysRevLett.106.020406} {\bibfield
  {journal} {\bibinfo  {journal} {Physical Review Letters}\ }\textbf {\bibinfo
  {volume} {106}},\ \bibinfo {pages} {020406} (\bibinfo {year}
  {2011})}\BibitemShut {NoStop}%
\bibitem [{\citenamefont {{Mauro D'Ariano}}\ \emph {et~al.}(2003)\citenamefont
  {{Mauro D'Ariano}}, \citenamefont {{Paris}},\ and\ \citenamefont
  {{Sacchi}}}]{tomography}%
  \BibitemOpen
  \bibfield  {author} {\bibinfo {author} {\bibfnamefont {G.}~\bibnamefont
  {{Mauro D'Ariano}}}, \bibinfo {author} {\bibfnamefont {M.~G.~A.}\
  \bibnamefont {{Paris}}}, \ and\ \bibinfo {author} {\bibfnamefont {M.~F.}\
  \bibnamefont {{Sacchi}}},\ }\href@noop {} {\bibfield  {journal} {\bibinfo
  {journal} {arXiv:quant-ph/0302028}\ } (\bibinfo {year} {2003})}\BibitemShut
  {NoStop}%
\bibitem [{\citenamefont {Carmeli}\ \emph {et~al.}(2012)\citenamefont
  {Carmeli}, \citenamefont {Heinosaari},\ and\ \citenamefont
  {Toigo}}]{informationally}%
  \BibitemOpen
  \bibfield  {author} {\bibinfo {author} {\bibfnamefont {C.}~\bibnamefont
  {Carmeli}}, \bibinfo {author} {\bibfnamefont {T.}~\bibnamefont {Heinosaari}},
  \ and\ \bibinfo {author} {\bibfnamefont {A.}~\bibnamefont {Toigo}},\ }\href
  {\doibase 10.1103/PhysRevA.85.012109} {\bibfield  {journal} {\bibinfo
  {journal} {Phys. Rev. A}\ }\textbf {\bibinfo {volume} {85}},\ \bibinfo
  {pages} {012109} (\bibinfo {year} {2012})}\BibitemShut {NoStop}%
\bibitem [{\citenamefont {Fano}(1957)}]{quorum}%
  \BibitemOpen
  \bibfield  {author} {\bibinfo {author} {\bibfnamefont {U.}~\bibnamefont
  {Fano}},\ }\href {\doibase 10.1103/RevModPhys.29.74} {\bibfield  {journal}
  {\bibinfo  {journal} {Rev. Mod. Phys.}\ }\textbf {\bibinfo {volume} {29}},\
  \bibinfo {pages} {74} (\bibinfo {year} {1957})}\BibitemShut {NoStop}%
\bibitem [{\citenamefont {Busch}\ \emph {et~al.}(2013)\citenamefont {Busch},
  \citenamefont {Lahti},\ and\ \citenamefont
  {Mittelstaedt}}]{busch2013quantum}%
  \BibitemOpen
  \bibfield  {author} {\bibinfo {author} {\bibfnamefont {P.}~\bibnamefont
  {Busch}}, \bibinfo {author} {\bibfnamefont {P.}~\bibnamefont {Lahti}}, \ and\
  \bibinfo {author} {\bibfnamefont {P.}~\bibnamefont {Mittelstaedt}},\ }\href
  {https://books.google.de/books?id=jInnCAAAQBAJ} {\emph {\bibinfo {title} {The
  Quantum Theory of Measurement}}},\ Lecture Notes in Physics Monographs\
  (\bibinfo  {publisher} {Springer Berlin Heidelberg},\ \bibinfo {year}
  {2013})\BibitemShut {NoStop}%
\bibitem [{\citenamefont {Gamel}(2016)}]{gamel}%
  \BibitemOpen
  \bibfield  {author} {\bibinfo {author} {\bibfnamefont {O.}~\bibnamefont
  {Gamel}},\ }\href {\doibase 10.1103/PhysRevA.93.062320} {\bibfield  {journal}
  {\bibinfo  {journal} {Phys. Rev. A}\ }\textbf {\bibinfo {volume} {93}},\
  \bibinfo {pages} {062320} (\bibinfo {year} {2016})}\BibitemShut {NoStop}%
\end{thebibliography}%
